\documentclass[twocolumn,trackchanges,twocolappendix]{aastex701}

\usepackage{amsmath}
\usepackage{graphicx}
\usepackage{float}
\usepackage{caption}
\usepackage{subcaption}
\usepackage{longtable}
\usepackage{graphicx}
\usepackage{xcolor}
\usepackage{orcidlink}

\begin{document} 
    \title{Improved reconstruction of the century-long solar magnetic field by incorporating morphological asymmetry in sunspots}

\author{Subhadip Pal\,\orcidlink{0009-0007-5204-3618}}
\affiliation{Department of Physics, Indian Institute of Technology Kanpur, Kanpur 208016, India}
\email{psubhadip24@iitk.ac.in}

\author{Gopal Hazra\,\orcidlink{0000-0001-5388-1233}} 
\affiliation{Department of Physics, Indian Institute of Technology Kanpur, Kanpur 208016, India}
\email{hazra@iitk.ac.in}

\author{Sudip Mandal\,\orcidlink{0000-0002-7762-5629}}
\affiliation{Max Planck Institute for Solar System Research, Justus-von-Liebig-Weg 3, 37077 Göttingen, Germany}
\email{mandals@mps.mpg.de}

  \begin{abstract} 
   Accurately modeling the solar magnetic field is important for understanding long-term solar activity and space weather, but it is challenging due to limited observations, especially near the Sun’s poles. The Surface Flux Transport (SFT) model simulates how magnetic flux moves across the solar surface and contributes to the polar field, but it parametrizes emerged sunspots as simple symmetric bipolar regions and needs improvement by including more realistic sunspot features. In this study, we reconstruct the century-long evolution of the Sun’s magnetic field, including the polar regions, using an improved SFT model. We incorporate cycle-dependent morphological asymmetry between leading and following sunspots, along with observationally derived tilt angles and sunspot area data for a century (1913–2016), to better represent magnetic flux transport and investigate the impact of asymmetry on polar field development. To study morphological asymmetry, we consider two cases: first, a long-term asymmetry factor calculated from the ratio of leading and following sunspots area spanning a century; second, the temporal asymmetry factor observed during solar cycle 23 applied to every solar cycle. Our simulated magnetic flux transport with inclusion of morphological asymmetry for both cases gets improved compared to the no asymmetry case in terms of enhanced low and mid-latitude magnetic flux and matches closely with observations. The simulated polar fields with asymmetry also show a better agreement with polar field observations for most cycles, particularly in capturing the timing of the polar field reversals and the peak amplitude during solar minima, which has severe consequences in solar cycle prediction.

    \end{abstract}  


%

    \section{Introduction}
    Understanding the solar magnetic field and its long-term evolution is of utmost importance for the present space-based world. The solar magnetic field decides long-term solar activity and its impact on the heliosphere and space weather. As we now understand, a key component of the solar dynamo \citep[e.g.,][]{Hazra2021}, the theory that explains the evolution and cycles of the solar magnetic field, is the strength of the polar magnetic field. Direct observation of this polar field has been challenging, primarily due to the constraints caused by projection effects. As a result, various alternative methods have been explored over the years.
    Historically, Surface Flux Transport (SFT) models \citep[e.g.,][]{Leighton64, Wang1991, Baumann_2004, Yeates2023, PN2025} have been very successful in simulating the evolution of magnetic flux on the solar surface and constraining the polar field. In most of the SFT models, usually, the active regions have been modeled as simple symmetric bipolar magnetic regions (BMRs) and incorporated as source terms in the magnetic flux transport equations to simulate their transport to the pole; until recently, some models have incorporated the observed magnetogram in the model \citep{Upton2014, Yeates2020}. For example, \citet{Yeates2020} found that a simple symmetric bipolar approximation of active regions leads to $24\%$ overestimation of the net axial dipole moment compared to the real magnetogram data of these regions, while keeping all other parameters unchanged. However, to reconstruct century-long flux transport of the solar magnetic field  \citep{Jiang_2011, Virtanen2022}, SFT models had to rely on the simple symmetric bipolar approximation of active regions, as no detailed magnetogram observations are available for the last century.
    
    Another realistic way to model the active regions is by considering the observed morphological asymmetry in the sunspot pairs \citep{Tlatov2014, TLATOV2015835, Murakozy2014, Iijima_2019}. Several observational studies (e.g., \cite{Tlatov2014}, \cite{Murakozy2014}) demonstrated the presence of a morphological asymmetry between leading and following sunspots. The morphological asymmetry is also found in theoretical models of flux emergence simulation of sunspots \citep{Fisher2000,Fan2009}. In the SFT modeling context, \citet{Iijima_2019} investigated this asymmetry effect using a one-dimensional (1D) SFT model. They have included an asymmetry in their Gaussian patches while modeling the BMRs, making the leading polarity sunspot narrower compared to the following sunspot. With the introduction of asymmetry, especially in the large and high-latitude BMRs, a significant reduction in the polar magnetic field has been observed in their model and the results are closer to the observation. \citet{Wang2021} studied in detail the effect of this asymmetry in terms of different strengths and complexities on the axial dipole moment that is another entity to quantify the polar field. They have considered the following sunspot to be more diffuse than the leading polarity and considered two cases with a strong and a weak asymmetry factor similar to \citet{Iijima_2019}. In addition, they have also considered a complex AR region. All three cases show a reduced contribution to the axial dipole moment compared to the symmetric BMR case (see Figure 4 of their paper). All of the above studies show clearly that morphological asymmetry in sunspot pairs, particularly the size asymmetry between leading and following sunspots, can significantly influence polar field formation by affecting cross-equatorial flux transport.

    In this paper, we present a methodology to reconstruct a century-long solar magnetic field by an improved approximation of bipolar active regions following \citet{Iijima_2019}. Instead of a simple symmetric bipolar approximation, we include morphological asymmetry in the sunspot pairs for the realistic depiction of the active region. Unlike previous studies \citep[e.g.,][]{Iijima_2019,Wang2021} that include a constant asymmetry factor over time, we have considered an observation-based time-dependent asymmetry factor. Also, \citet{Iijima_2019} and \citet{Wang2021} have not performed any simulations to reconstruct the century-long solar magnetic field. The studies that simulate the long-term reconstruction of solar magnetic field using surface flux transport (SFT) simulations (e.g., \cite{Jiang_2011}, \cite{Hofer_2024}) did not account for the effects of morphological asymmetry. Hence, for the first time, we incorporate a time-dependent morphological asymmetry factor to all solar cycles to simulate realistically the century-long temporal evolution of the photospheric magnetic field and polar field. This enables us to have an evaluation of the improvement on the simulation accuracy achieved by including time-dependent morphological asymmetry while modeling active regions.
    
    The plan of the paper is as follows. In the next section, we briefly describe our SFT model and explain how we include morphological asymmetry for modeling active regions. In section 3, we describe the observational asymmetry parameters that we consider for all cycles over the last century. We present our results with morphological asymmetry in section 4. We also discuss how morphological asymmetry affects the overall century-long magnetic field reconstruction and polar field compared to the simple symmetric bipolar approximation in this section. Finally, we conclude our findings in section 5.
    \section{SFT Model with morphological asymmetry in sunspot pairs}\label{sec:model}
    To reconstruct a century-long detailed solar magnetic field on the surface of the Sun, we adopt a two-dimensional traditional SFT model \citep{Baumann_2004, Cameron_2010, Yeates2015}. The SFT model computes passive transport of the radial component of the magnetic field, influenced by differential rotation, meridional flow, and super-granular diffusion. The governing equation is given below:
        \begin{align}
            \frac{\partial B}{\partial t}=&-\Omega(\lambda)\frac{\partial B}{\partial \phi}-\frac{1}{R_{\odot}\cos{\lambda}}\frac{\partial}{\partial \lambda}[v(\lambda)B \cos{\lambda}]& \notag \\ 
           &+\eta_H\left[\frac{1}{R_{\odot}^2 \cos{\lambda}} \frac{\partial}{\partial \lambda}\left( \cos{\lambda} \frac{\partial B}{\partial \lambda}\right) + \frac{1}{R_{\odot}\cos^2{\lambda}}\frac{\partial^2 B}{\partial \phi^2} \right]& \notag \\ 
           &+S(\lambda,\phi,t)+D(\eta_r),
           \label{Induction Equation}
       \end{align}
       where $B(\equiv B_r)$ is the radial component of the magnetic field on the surface of the Sun. $S(\lambda,\phi,t)$ represents the source term for magnetic flux, characterizing the emergence of bipolar magnetic regions at latitude $\lambda$, longitude $\phi$, and time $t$. $D(\eta_r)$ is the decay term due to the radial decay (for our case, we assume $D(\eta_r) = - \frac{B}{\tau}$ where { $\tau$ is a time scale of 10.2 years}), and the super-granular diffusion coefficient is $\eta_H = 250 \ \text{km}^2/\text{s}$. The differential rotation ($\Omega(\lambda)$) and meridional flows ($v(\lambda)$) are taken as observed by using analytical formulas. The differential rotation and meridional flow profile are taken from \citet{Snodgrass_1983} and  \citet{Whitbread_2017} respectively and given below:
    \begin{eqnarray}
       \Omega(\lambda)=13.38-2.30\sin^2{\lambda}-1.62\sin^4{\lambda}  \label{eq:diff_rot}\\
        v(\lambda) = R_\odot \Delta_v\sin(\lambda)\cos^{2.33}(\lambda)
    \end{eqnarray}
    where $\Delta_v = 0.4 \times 10^{-7}$. These flow profiles are widely used in many SFT simulations to date \citep[e.g.,][]{Whitbread_2017}. After providing the velocity profile and diffusivity, our main task is to model the source function $S(\lambda, \phi, t)$ appropriately to account for the emerging active regions on the surface of the Sun. Once the source function is defined, we solve the above equation following an approach given by \citet{Yeates2015}, where the magnetic field is written in terms of magnetic vector potentials and solved in the spherical polar coordinates. Next, we explain how we model the source function $S(\lambda, \phi, t)$ for both a simple symmetric bipolar approximation and by accounting for morphological asymmetry in sunspot pairs. 
 
    \subsection{Modeling active regions as symmetric bipolar regions}
    Active regions are generally closely associated with two areas of opposite magnetic polarity, typically observed as pairs of sunspots (hence bipolar magnetic regions (BMRs)) BMRs are the primary source of the large-scale magnetic field on the Sun's surface. They can vary widely in size and shape. One of the main difficulties while reconstructing the evolution of solar magnetic fields over the last hundred years is the lack of data about the size \& shape, distribution \& polarity of the magnetic field in those spots. Thanks to historical observations \citep{Fligge1997, Hathaway2002, Vaquero2007, Mandal2017, Mandal2020}, we have the total sunspot area measured from many observatories, including their emergence latitude, longitude and time. In the classic SFT model \citep[e.g.,][]{Ballegooijen_1998, Cameron_2010, Jiang_2011, Jiang2014b, Yeates2023}, these BMRs are modelled as circular magnetic patches. The magnetic field in a BMR consists of two opposite-polarity sunspots
    $B(\lambda,\phi)=s_LB^L(\lambda,\phi)+s_FB^F(\lambda,\phi)$, where L and F stand for the leading and following sunspot, and $s_L$ and $s_F$ are the signs of these two polarities, respectively. The magnetic field is defined for those polarities as 
    \begin{equation}
    B^M(\lambda,\phi)= B_{max}^M \exp\left({-\frac{2[1-\cos{\beta^M(\lambda,\phi)}]}{(\delta^M)^2}}\right),
    \label{eq: BMR_field}
    \end{equation} 
    where M can be L or F. $B_{max}$ is the maximum field within the patch. $\delta^M$ is the spatial size of the patches in radians. $\beta^M$  is the heliocentric angle between the location of the polarities and their centers and can be written as 
    \begin{equation}
    \cos\beta^M=\sin\lambda\sin\lambda^M+\cos\lambda\cos\lambda^M\cos(\phi-\phi^M).
    \end{equation}
    $(\lambda^M,\phi^M)$ are the latitude and longitude of the center of the two polarities, respectively. These can be calculated from the position of the total sunspot like:
        \begin{align}
            \lambda^L=\lambda^c-0.5s_\lambda\Delta\beta\sin\alpha \\
            \lambda^F=\lambda^c+0.5s_\lambda\Delta\beta\sin\alpha \\
            \phi^L=\phi^c+0.5\Delta\beta\cos\alpha(\cos\lambda^c)^{-1} \\
            \phi^F=\phi^c-0.5\Delta\beta\cos\alpha(\cos\lambda^c)^{-1} 
        \end{align}
    $(\lambda^c,\phi^c)$ are the central latitude and longitude of the total sunspot. $\alpha$ is the tilt angle of the BMR. $\Delta\beta$ is the angular distance between the centers of the two polarities.
    Previous studies \citep[][]{Ballegooijen_1998, Cameron_2010, Iijima_2019} have constrained this value to be approximately $\Delta\beta=\delta^L/0.4$, where $\delta^L$ is the patch size of the leading polarity in radian.
    
    \subsection{Modeling active regions with morphological asymmetry}
\begin{figure}[H]
            \centering
           {(a) HMI$_{\mathrm{cont}}$}\par
            \includegraphics[width=1\linewidth]{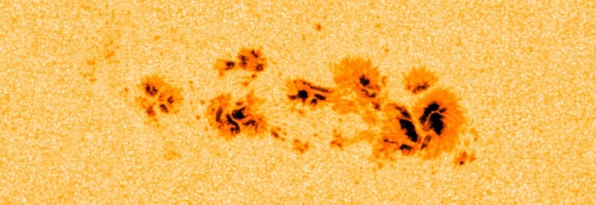}
           \vspace{0.1cm} 
           {(b) HMI$_{\mathrm{LOS}}$ magnetogram }\par
          \includegraphics[width=1\linewidth]{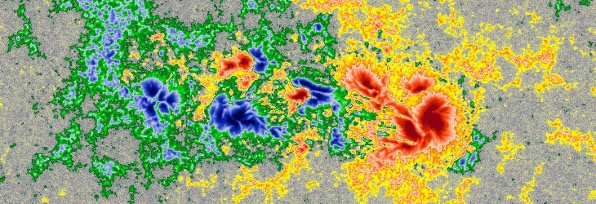}
         \caption{A representative example of the morphology and magnetic configuration of a typical sunspot group. Panel-a displays a white-light image of the sunspots captured by the Helioseismic and Magnetic Imager (HMI) 4500~\AA~ continuum channel, while Panel-b shows their photospheric line-of-sight (LOS) magnetic configuration as recorded by the HMI 6173~\AA~ channel. In the magnetogram, the yellow and red colors represent the north (outward) polarity, whereas the green and blue colors indicate the south (inward) polarity. Credit: SDO/HMI \citep{2012SoPh..275..207S}}
           \label{sunspot picture}
    \end{figure}

    The morphological asymmetry of sunspot pairs is expected to play a crucial role in the formation of the polar field. After their emergence, the following sunspot tends to become more diffused than the leading sunspot (\citealp{Fisher2000,Fan2009}). As shown in Figure~\ref{sunspot picture}, sunspots are inherently asymmetric. The following spots appear more scattered and diffuse compared to the leading ones, 
    thereby allowing the following spots to contribute significantly to the polar field. Moreover, due to its larger size, it undergoes cross-equatorial transport, influencing the leading sunspot in the opposite hemisphere and thereby affecting the polar field (see Figure~1 of \citealp{Iijima_2019}).

    In previous sections, we discussed the placement of magnetic fields in the sunspot region using a Gaussian circular patch. The flux corresponding to a sunspot region is given by:
        \begin{equation}
            \Phi^M \sim \pi R_{\odot}^2(\delta^M)^2B_{max}^M
            \label{flux}
        \end{equation}
        where $\delta^M$ is the patch size. Therefore, the total flux from a bipolar magnetic region (BMR) is:
        \begin{equation}
            \Phi_{tot}=\Phi^L+\Phi^F
            \label{flux_balance}
        \end{equation}
        To ensure flux balance and prevent monopole generation, the fluxes between the two polarities must satisfy the relation:
        \begin{equation}
            B_{max}^L(\delta^L)^2=B_{max}^F(\delta^F)^2
        \end{equation}
        We assume a threshold magnetic field ($B_T$) at the edge of the sunspot, i.e., at $r = \delta_{spot}^M$. Considering the sunspots as circular patches with radius $r$, the threshold field is expressed as:
        \begin{equation}
            B_T=B_{max}^M\exp({-\delta^M_{spot}}/\delta^M)
            \label{b_t}
        \end{equation}
        where $\delta^M_{\text{spot}}$ is the sunspot radius.
        From Equation~(\ref{flux}), the sunspot area can be written as:
        \begin{equation}
            A_{spot}^M=\pi R_{\odot}^2(\delta^M_{spot})^2
            \label{area}
        \end{equation}
        Using Equations~(\ref{flux}), (\ref{b_t}), and (\ref{area}), we derive the following relation:
        \begin{equation}
            B_T=B_{max}^M\exp({-A_{spot}^MB_{max}^M}/\phi^M)
        \end{equation}
        Given that $B_T$ is known, $B_{\text{max}}^M$ can be expressed in terms of $B_T$ as:
        \begin{equation}
            B_{max}^M=\frac{-\Phi^M}{A_{spot}^M}W_0 \left(-\frac{B_TA_{spot}^M}{\Phi^M}\right)
            \label{bmax}
        \end{equation}
        where $W_0(x)$ is the principal branch of the Lambert W function. The patch size can then be expressed as:
        \begin{equation}
            \delta^M=\delta_{spot}^M\left[-W_0 \left(-\frac{B_TA_{spot}^M}{\Phi^M}\right)\right]^{-0.5}
            \label{delta}
        \end{equation}
        Detailed calculations can be found in \cite{Iijima_2019}, which provides a relationship between the threshold magnetic field ($B_T$), the total flux from the BMR ($\Phi_{tot}$), and the BMR area ($A_{spot}$):
        \begin{equation}
            \Phi_{tot}=cB_TA_{spot}
            \label{flux_with_para}
        \end{equation}
        In our simulations, we use $c = \frac{2e}{1 + \text{f}_{\text{spot}}}$, where f$_{\text{spot}}$ is the asymmetry parameter defined as the ratio of the following to the leading sunspot areas:
        \begin{equation}
            \text{f}_{\text{spot}}=\frac{A_{spot}^F}{A_{spot}^L}
            \label{asymmetry parameter}
        \end{equation}
        Since the total sunspot area $A_{spot} = A_{spot}^L + A_{spot}^F$, Equation~(\ref{asymmetry parameter}) allows us to express the leading and following sunspot areas as:
        \begin{align}
            A_{spot}^L=\frac{A_{spot}}{1+\text{f}_{\text{spot}}} \\
            A_{spot}^F=\frac{A_{spot}\text{f}_{\text{spot}}}{1+\text{f}_{\text{spot}}}
        \end{align}
        In accordance with the flux balance condition between the two polarities, it follows from Equation~(\ref{flux_balance}) that the fluxes can be expressed as 
        \begin{equation}
            \Phi^L=\Phi^F = \Phi_{tot}/2
            \label{eq29}
        \end{equation}
        So, from Equations~(\ref{bmax}), (\ref{delta}), (\ref{flux_with_para}) and (\ref{eq29}) it can be written as:
        \begin{align}
            B_{max}^L &= \frac{-\Phi_{tot}/2}{\frac{A_{spot}}{1+\text{f}_{\text{spot}}}}W_0 \left(-\frac{B_T\frac{A_{spot}}{1+\text{f}_{\text{spot}}}}{\Phi_{tot}/2}\right) \notag \\
            &=-\frac{cB_T(1+\text{f}_{\text{spot}})}{2}W_0\left(-\frac{1}{c} \frac{2}{1+\text{f}_{\text{spot}}} \right) \notag \\
            \implies &B_{max}^L=-eB_TW_0(-\frac{1}{e})=eB_T
        \end{align}
        Similarly,
        \begin{align}
            B_{max}^F &= -\frac{eB_T}{\text{f}_{\text{spot}}}W_0(-\frac{\text{f}_{\text{spot}}}{e}), {\rm and}
        \end{align}
        \begin{align}
            \delta^L &=\delta_{spot}^L\left[ -W_0\left(-\frac{1}{e} \right)\right]^{-0.5} = \delta_{spot}^L
            \label{leading_patch}
        \end{align}
        \begin{align}
            \delta^F &=\delta_{spot}^F\left[ -W_0\left(-\frac{\text{f}_{\text{spot}}}{e} \right)\right]^{-0.5}
            \label{following_patch}
        \end{align}
    The patch sizes for leading and following polarity due to asymmetry factor f$_{\rm spot}$ are calculated from Equations~(\ref{leading_patch}) and (\ref{following_patch}) and incorporated in Equation~(\ref{eq: BMR_field}) to estimate the source function $S(\lambda,\phi, t)$ in the SFT equation.

    \subsection{Tilt angle} \label{sec:tilt}
    In the SFT simulations, the tilt angle significantly influences the polar field \citep{Baumann_2004} and its accurate measurement is essential for simulating a reliable polar field.  According to Joy's law, the tilt angle is proportional to the emergence latitude $\lambda$ of the BMRs, but observationally a scatter around the tilt has been observed \citep{Wang_2015, Jiao_2021}. Here we have followed the approach of \citet{Cameron_2010} to calculate the tilt angle as given below:
    \begin{equation}
        \alpha = g_{inflow}T_n^{sqr}\sqrt{\lambda} + \epsilon
    \end{equation}
    where $g_{inflow}=0.7$ and $\epsilon$ is the tilt angle scattering \citep{Jiang_2011, Jiao_2021}. The factor $T_n^{sqr}$ is cycle dependent Tilt angle co-efficient to fit the tilt angle data with Joy's law. For all of our simulations, we have considered no tilt angel scattering and hence we put $\epsilon = 0$ in our simulation. 
    
    \citet{Jiao_2021} conducted a comparative study of sunspot group tilt angle coefficients ($T_n^{sqr}$) from three major observatories: Kodaikanal Solar Observatory (KSO), Mount Wilson Observatory (MWO), and the Debrecen Photoheliographic Data (DPD). For earlier solar cycles (Cycles 15 to 21), data were available from both MWO and KSO, while for later cycles (Cycles 21 to 24), DPD provided the relevant tilt angle information.

    By analyzing the overlap in Cycle 21, \citet{Jiao_2021} found that MWO data showed a better agreement with the DPD dataset compared to KSO. However, they also identified inconsistencies in the MWO tilt angle data for Cycles 15 and 19, suggesting possible reliability issues in those periods.
    In our study, we use the best fitted $T_n^{sqr}$ reported in literature so far to maintain physical realism in the Surface Flux Transport (SFT) simulations.
    When MWO data were used directly, significant discrepancies appeared in the simulated polar field during Cycle 19 as already reported. To ensure consistency, we chose to use the Kodaikanal (KSO) tilt angle data for Cycles 15 through 20. For Cycles 21 to 22, we utilized the unbinned DPD tilt angle co-efficient data (see Table \ref{Tilt_table} ).

    We have not used the DPD tilt angle data for Cycles 23 and 24, as \citet{Jiang2013} highlighted that statistically derived $T_n^{sqr}$ for these cycles can lead to an overestimation of the polar field in SFT simulations compared to observations. Instead, for Cycles 23 and 24 \footnote{\textcolor{red}{*} The tilt angle coefficients for Cycles 23 and 24 are 1.1 and 1.9, respectively, when asymmetry is not considered, and 1.3 and 1.6, respectively, when asymmetry is included.}, we used the tilt angle data provided by \citet{Iijima_2019}, which are considered more suitable for SFT simulations during this period. Note that the $T_n^{sqr}$s are different for Cycle 23 and Cycle 24 for both symmetry and asymmetry cases. 

    In our simulation, we have not imposed any initial polar field. Instead, we initialize the Surface Flux Transport model from the year 1874 and allow the model to self-consistently build up the polar field over time. We then consider the simulated polar field values starting from 1913 onward. For solar cycles preceding Cycle 15, where tilt angle data are not available and we estimate the tilt angle coefficient using the empirical relation $T_n = -0.0021 S_n + 1.72$ (\citealp{Hofer_2024}), where $T_n$ is the tilt angle coefficient and $S_n$ is the sunspot number of the corresponding cycle.

    \begin{table}
        \centering
        \caption{Tilt angle coefficient derived from observational data of the Kodaikanal Solar Observatory (Cycles 15–20) and the Debrecen Photoheliographic Database (Cycles 21–24)}
        \label{Tilt_table}
        \setlength{\tabcolsep}{3.0pt}  
        \begin{tabular}{|c|c|c|c|c|}
            \hline
            Data & Cycle 15 & Cycle 16 & Cycle 17 & Cycle 18  \\
            \hline
            $T_n^{sqr}$($\Delta \beta \geq 2.5^\circ$) & 1.83   & 1.60  & 1.56 & 1.61  \\
            \hline
            $T_n^{sqr}$($\Delta \beta \geq0^\circ$)  & 1.48  & 1.45   & 1.47 & 1.17  \\
            \hline
            \hline
            Data & Cycle 19 & Cycle 20 & --- & ---\\
             \hline
            $T_n^{sqr}$($\Delta \beta \geq 2.5^\circ$) & 1.35   & 1.70  & --- & ---   \\
            \hline
            $T_n^{sqr}$($\Delta \beta \geq0^\circ$)  & 1.11  & 1.34 & --- & ---   \\
            \hline
        \end{tabular}
    
        \hspace{0.5 cm}
        \setlength{\tabcolsep}{1.5pt}  
        \begin{tabular}{|c|c|c|c|c|}
            \hline
            Data & Cycle 21 & Cycle 22 & Cycle 23${^{\textcolor{red}{*}}}$ & Cycle 24${^{\textcolor{red}{*}}}$ \\
            \hline
            $T_n^{sqr}N$($\Delta \beta \geq 2.5^\circ$) & 1.81 & 1.66 & 1.96 & 1.80 \\
            \hline
            $T_n^{sqr}N$($\Delta \beta \geq0^\circ$)  & 1.55  & 1.59 & 1.79 & 1.66  \\
            \hline
            $T_n^{sqr}S$($\Delta \beta \geq 2.5^\circ$) & 1.54 & 1.67 & 1.87 & 2.07 \\
            \hline
            $T_n^{sqr}S$($\Delta \beta \geq0^\circ$)  & 1.41  & 1.45 & 1.71 & 1.84  \\
            \hline
        \end{tabular}
    
    \end{table}
    
    \section{Determination of the Asymmetry Parameter from observation}\label{sec:asymmetry}
    Observationally, the morphological asymmetry between leading and following sunspots is well established \citep{Grotrian1950, Zwaan1981, Murakozy2014, Tlatov2014, TLATOV2015835}. The asymmetry appears in shape, size, magnetic field and spatial separation between leading and following sunspots and quantifying them over a century-long time period is not a trivial task. By employing an automated image processing algorithm on white-light images at the Kislovodsk Solar Station, \citet{Tlatov2014} found that leading polarity sunspot areas are on average approximately 2.5 times the area of the following sunspots for cycle 23. Given the fact that sunspot locations and sizes change with latitude and phase of the solar cycle, the asymmetry parameters are naturally expected to vary with time and latitude. In fact, for cycle 23, \citet{Tlatov2014} indeed reported a time-dependent asymmetry including hemispheric asymmetry (see Figure~5 of their paper). Considering this fact, we have incorporated a time-dependent asymmetry profile into our simulations. Unlike \cite{Iijima_2019}, who utilized a cycle-averaged constant asymmetry factor of $\text{f}_{\text{spot}}=0.4$ based on the work of \cite{Tlatov2014}, we account for its temporal variation as well. 

\begin{figure}[!htbp]
        \centering
        \includegraphics[width=0.95\linewidth]{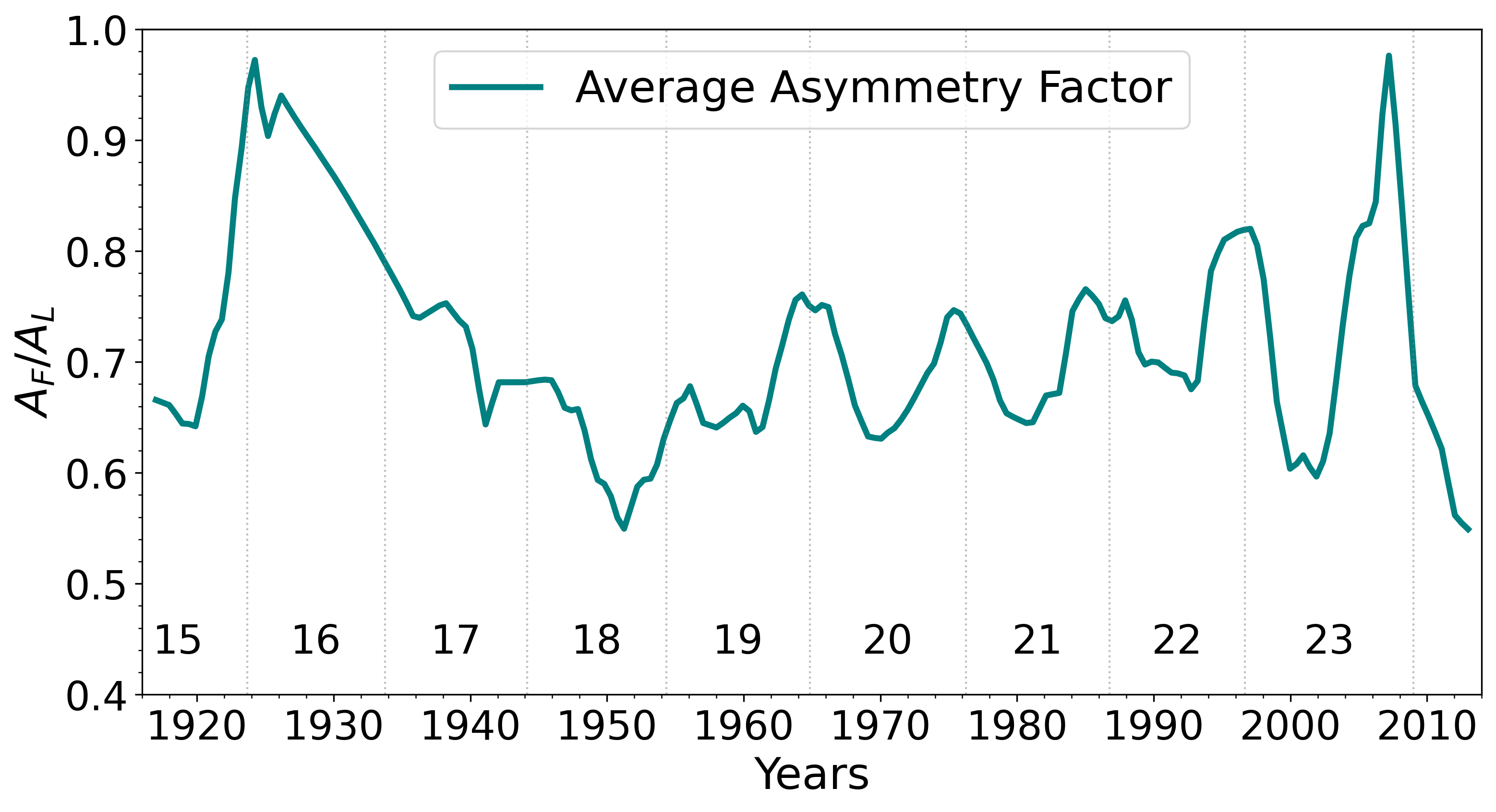}
        \caption{Ratio of the 5-year running mean areas of umbra and pores of Following ($A_F$) and Leading ($A_L$) polarity in units of $\mu hm$ based on \cite{TLATOV2015835}.}
            \label{fig:long_term_fspot}
    \end{figure}

    To extract the time-dependent values of the asymmetry factor ($\text{f}_{\text{spot}}$) from observations, we utilized the dataset provided by \cite{TLATOV2015835}, which uses information from Mount Wilson Observatory (MWO). We analyzed this catalog covering the period from 1917 to 2013 by calculating the ratio of the 5-year running mean areas of umbra and pores for both the following ($A_F$) and leading ($A_L$) polarities. The derived asymmetry parameter $\text{f}_{\text{spot}} = A_F/A_L$ is shown in Figure~\ref{fig:long_term_fspot}. Although asymmetry parameter data is available from 1917 to 2013, our simulation spans a longer period—from 1913 to 2016. For the years where data for $A_F$ and $A_L$ are not available from MWO, we use a fixed asymmetry factor of 0.8 uniformly across those cycles in the model based on \cite{Murakozy2014}.
    So far, this approach provides the best realistic representation of the long-term temporal behavior of the asymmetry factor over the past century.

\section{Reconstruction of solar magnetic field over a century}
    
In this study, we aim to successfully simulate the surface transport of magnetic flux using our modified SFT model with included morphological asymmetry in sunspots for a century from 1913-2016. The three main parameters required to faithfully simulate the evolution of the magnetic field on the solar surface (i.e, photosphere) are the location \& area of sunspots, tilt angle of sunspot groups, and their asymmetry. A complete list of observationally constrained parameters are explained in section~\ref{sec:model}. 

The data for sunspot areas and emergence locations were taken from the catalog by \citet{Mandal2020}, which provides daily observations but does not include sunspot group numbers. To avoid counting the same sunspot group multiple times, we used the RGO dataset (\url{https://solarscience.msfc.nasa.gov/greenwch.shtml}) \citep{Hathaway_2003}, which includes sunspot group numbers along with position and area information. From this dataset, we selected each sunspot group only on the day it reached its maximum area \citep{Baumann_2004,Jiang_2009}, ensuring that each group is counted once. We then matched these filtered RGO entries with those in the catalog of \citet{Mandal2020}  by comparing the latitude, longitude, and time of emergence. If a sunspot group in the catalog closely matched one of the filtered RGO groups, we retained it. In this way, we ensured that we used the more reliable area measurements from the catalog (\citet{Mandal2020}) and have corrected for the repetition of sunspots in our model. Details about the tilt angles and, spot asymmetry parameters used for all of our simulations is described in section~\ref{sec:tilt} and ~\ref{sec:asymmetry}, respectively.

A total of three cases are considered in this study. In Case-1, we run our simulation without introducing any asymmetry in sunspots that is f$_{spot}=1$. In Case-2, we incorporate the long-term asymmetry in sunspots as shown in Figure~\ref{fig:long_term_fspot}. Finally, for Case-3, we consider the temporal asymmetry of cycle 23 as derived in \citet{Tlatov2014}, and applied the same asymmetry for all cycles so as to compare how a difference in the asymmetry parameter might change the century-long evolution of the magnetic field (see Appendix A for details about the estimated cycle 23 asymmetry that taken into our SFT simulation).

\begin{figure*}[!htbp]
        \centering
        \includegraphics[width=0.99\textwidth]{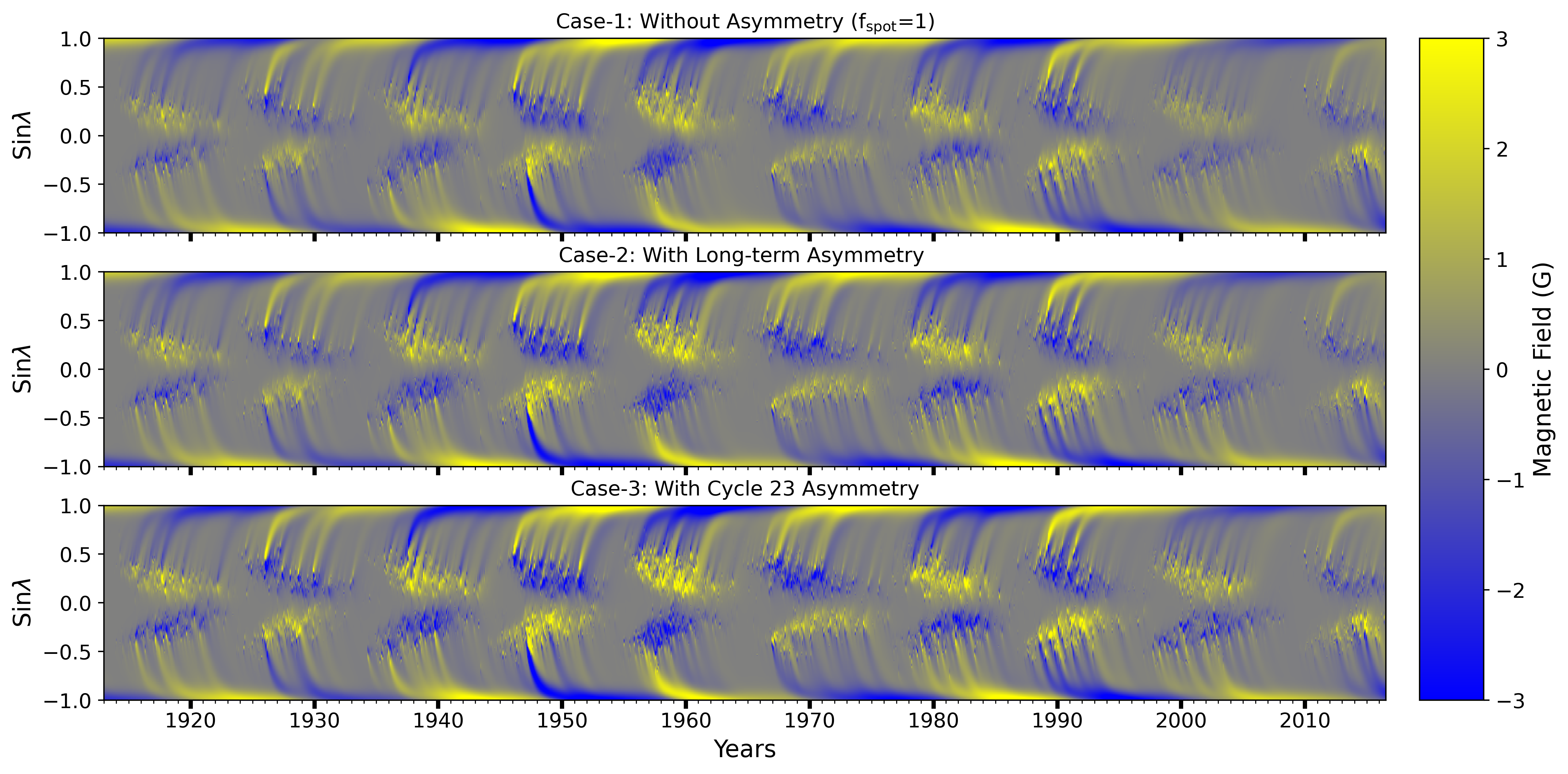} 
        \caption{Century-long evolution of photospheric magnetic field (Butterfly Diagram) for about 100 years (1913-2016) from our simulations. The upper panel shows the butterfly diagram without inclusion of morphological asymmetry (Case-1), the middle panel shows the butterfly diagram with long-term asymmetry variation in sunspots (Case-2), and the lower panel shows the butterfly diagram with inclusion of cycle-23 morphological asymmetry (Case-3) factor for all cycles.}
        \label{butterfly diagram}
    \end{figure*}

Figure~\ref{butterfly diagram} illustrates all three time-latitude diagrams (commonly referred to as `butterfly diagrams') from the three case runs of our SFT simulation.A direct comparison of Case-2 \& Case-3 (with morphological asymmetry) with Case-1 (without asymmetry) in Figure~\ref{butterfly diagram} makes it evident that the magnetic field for asymmetric cases in the low and mid-latitudes activity belt  (($\leq 55^{\circ}$)) is grainier than the case without asymmetry. The similar results were also reported in \citet{Iijima_2019}. The term {\it grainy} was introduced by \citet{Jiang_2014a} representing the enhancement of low and mid-latitudes magnetic fluxes compared to high latitudes. Magnetogram observations are not available for all the cycles that we considered here, but if we compare our results with available NSO magnetogram observations for the cycle 21-24 (see panel d of Figure~8 in \citealp{Iijima_2019}), we find that the magnetic flux transport for the asymmetric cases matches better with observations. If we do not consider asymmetry in sunspots (Case-1), the ratio of magnetic flux in the low and mid latitudes activity belt to high-latitude becomes lower than magnetogram observations, but that gets improved when we include the morphological asymmetry. The reason is straightforward to understand. For the asymmetric cases, the following polarity sunspot, being more diffused and distributed over a bigger area compared to the leading polarity sunspot, moves faster to the pole by meridional flow and diffusion. This quick movement of following polarity sunspot does not give enough time for the sunspot pairs to cancel out flux between each other, which usually happens for the symmetric case when both sunspots move to the pole almost at the same time. For Case-3, the asymmetry considered is larger than the long-term asymmetry considered in Case-2, resulting in a much grainier butterfly diagram than in Case-2.

    \subsection{Polar field evolution} \label{sec:polarfield}
    We have calculated the polar magnetic field from our simulation over the latitude range of $\pm 70^\circ$ to the pole for all cases without asymmetry (Case-1) and with asymmetry (Case-2 \& Case-3). The simulated polar fields for all cases with and without the inclusion of the morphological asymmetry factor are compared with observational data. A direct measurement of the polar field is available from 1976 to the present from Wilcox Solar Observatory; however, historical measurements of the polar field before 1976 are not available directly and different proxies have been used to reproduce the polar field for those past years. Recently, using Ca II K polar index data from Kodaikanal Solar Observatory (1904-2007) and Rome Precision Solar Photometric Telescope (2000-2022), \citet{Mishra2025} have provided measurements of polar field in the last century. Using the polar facule count from Mount Wilson Observatory \citet{Munoz2012} estimated the polar field as well. As the historical reconstruction of the polar field is model and proxy-dependent, we will be comparing our simulated polar field with both observationally constructed KSO \citep{Mishra2025} and MWO \citep{Munoz2012} polar field data. We note that for polar field observations, before 1976, we used the Ca II K polar index proxy data for KSO and polar facule data for MWO. After 1976, we merged the WSO polar field data into both the KSO and MWO datasets. 

\begin{figure*}[!htbp]
            \centering
            \includegraphics[width=0.98\textwidth]{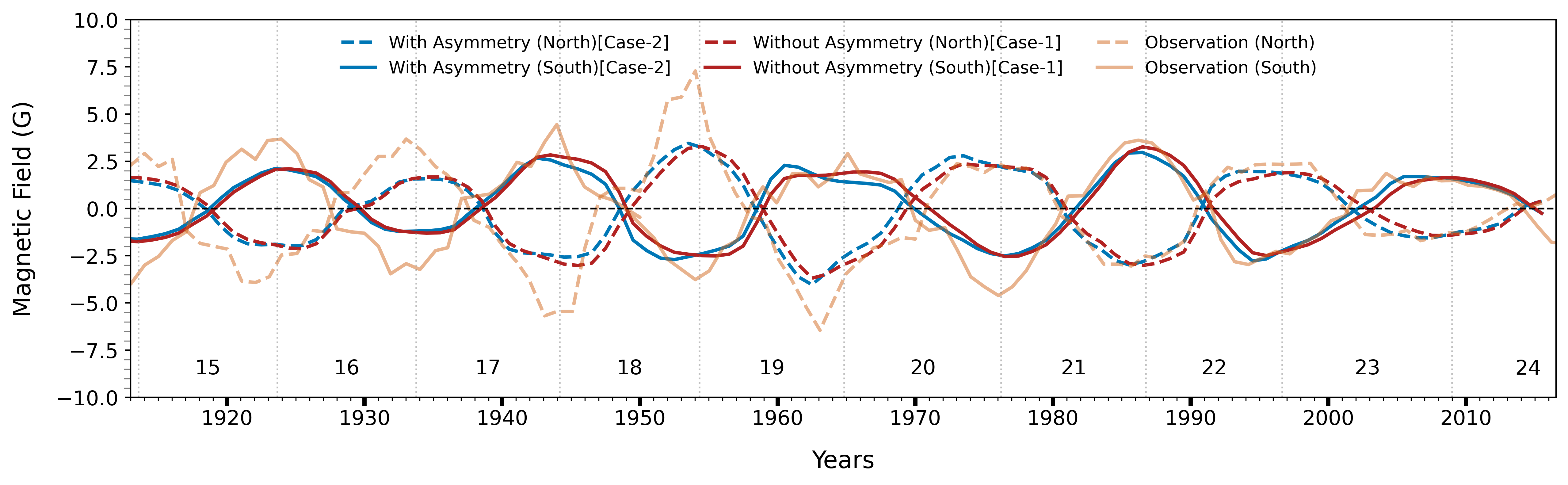} 
            \caption{Comparison of the polar field over 100 years from the simulation, with and without the long-term temporally varying asymmetry factor, against observational polar field data from Kodaikanal Solar Observatory (KSO). The solid line represents the south pole, while the dotted line represents the north pole. The observed polar field from WSO  is merged with KSO polar field data after 1976. The vertical grey dotted lines show the position of solar minima.}
            \label{polar_field_kso}
    \end{figure*}

    The comparison of the simulated polar field from our model with the observed polar field from Kodaikanal observatory is shown in Figure~\ref{polar_field_kso}. We only show here simulated polar fields for two cases - without asymmetry (Case-1) and long-term morphological asymmetry (Case-2). The yellow lines represent the observed polar field from \cite{Mishra2025}, while the red and cyan lines correspond to the polar field calculated from our simulation for Case-1 without asymmetry and for Case-2 with long-term asymmetry, respectively. The dotted and solid lines indicated the north and south poles, respectively, for all scenarios. Instead of specifying an initial value for the polar field, we started the simulation in 1874 to fix the initial offset of the polar field properly for both of our cases. This consideration makes our polar field data from 1906 consistent with observations. The simulated polar fields for both cases (Case-1 and Case-2) match qualitatively well with the observed polar field data. However, if we look carefully, it is clear that our simulated polar fields for both cases get improved for the cycles after 1976, as polar field observations available after 1976 are more reliable, with some differences between the two cases.  

\begin{figure*}[!htbp]
            \centering
            \includegraphics[width=0.98\textwidth]{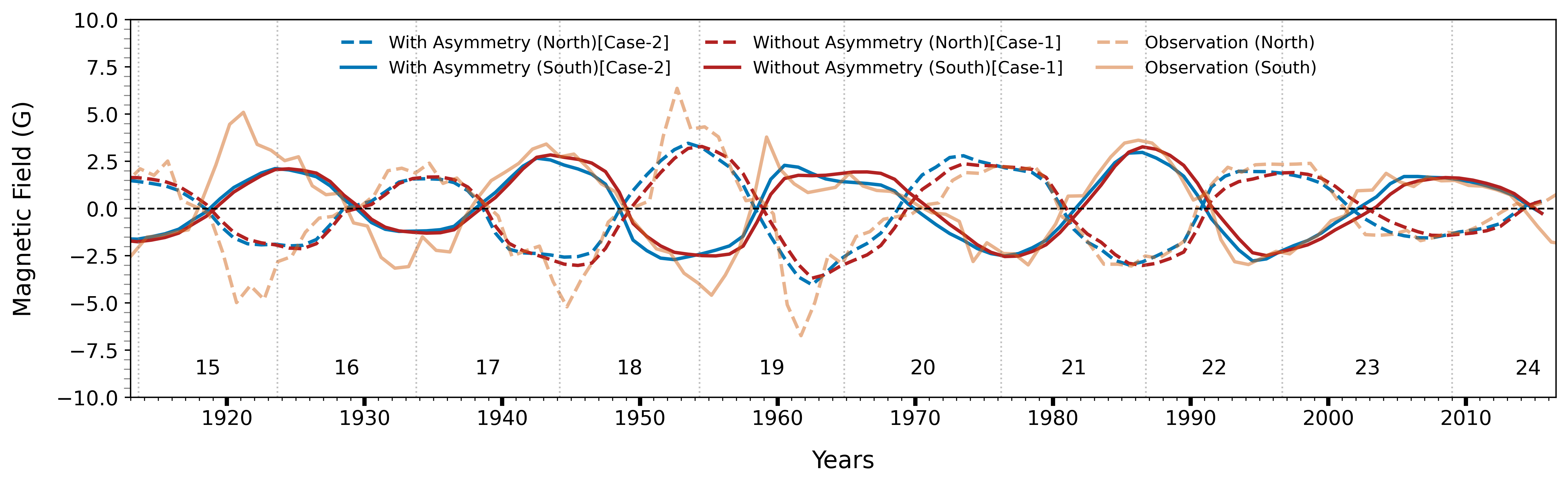} 
            \caption{Same as Figure~\ref{polar_field_kso} but with observational polar field data from MWO (1913-1976) and WSO (1976-2016).}
            \label{polar_field_mwo}
    \end{figure*}

    A similar comparison is also done with the MWO polar field data \citep{Munoz2012} as shown in Figure~\ref{polar_field_mwo}. In this case, we also see simulated polar fields for both cases match well with the observed data, although there are differences between two cases.  For Case-2 with long-term asymmetry, cycle reversal happens close to the observations and earlier than the Case-1, where no morphological asymmetry is considered, as diffuse following polarity sunspots quickly move to the pole and cancel the existing polar field.

    For our simulation, we used a magnetic field strength of $B_T=366 \ G$ (for Case-1) to match the observed polar field strength of \citet{Munoz2012}. While the amplitude is influenced by the free parameter $B_T$, we estimated $B_T$ in such a way that the $B_{max}$ is very close to the value used in some previous studies \citep{Cameron_2010} and maintaining consistency with MDI polar field data scale. To preserve the total magnetic flux while introducing asymmetry, as the area of the following polarity is increased, the magnetic field strength \( B_T \) is scaled according to the asymmetry factor \( \text{f}_\text{spot} \), ensuring flux conservation relative to the symmetric configuration. Also, from polar field comparison plots (Figure~\ref{polar_field_kso} and ~\ref{polar_field_mwo}), we see that for cycles 18-24, in the Case-2 (long-term asymmetry included), the polar field reversal happens before the Case-1 (without any asymmetry), but for earlier cycles 15-17, the differences in reversal time for both cases are not that significant. The peak amplitude of the polar field varies, but not very significantly, between the cases with or without asymmetry. It is expected that the different solar cycles will have different asymmetry factors, and hence we do not show here Case-3 in which we implemented the asymmetry variation of cycle 23 in all cycles. However, for quantitative comparison for polar field time reversal and peak amplitude with observational data, we show all three cases as explained in the next section.   

    \subsubsection{Polar field reversal time}
     As shown in Figure~\ref{polar_field_kso} and \ref{polar_field_mwo}, the polar field for both Case-1 and Case-2 qualitatively aligns with the observed polar field. However, significant differences were observed between the two cases. Specifically, the timing of the polar field reversal and the amplitude of the polar field differed when the morphological asymmetry factor of sunspots was included versus when it was omitted. Both of these quantities are crucial, as they dictate the timing and amplitude of the next cycle maximum. We present a thorough analysis of those differences. 

     \begin{figure*}[!htbp]
            \centering
            \textbf{~~~~~~~~~~~~KSO data \citep{Mishra2025}~~~~ ~~~~~~~~~~~~   ~~~~~~~~MWO data \citep{Munoz2012}}\par\medskip
            \includegraphics[width=0.49\linewidth]{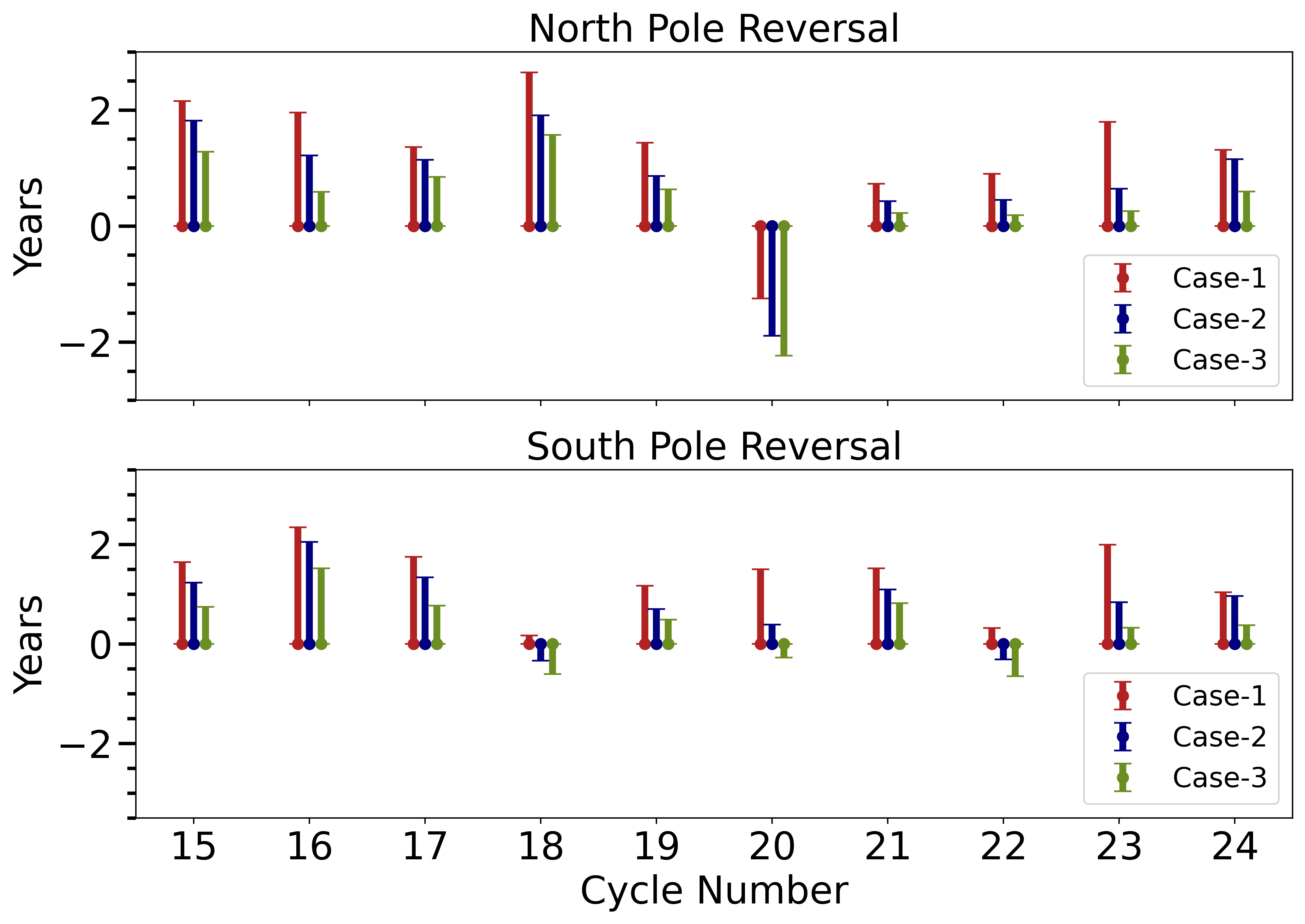}
            \includegraphics[width=0.49\linewidth]{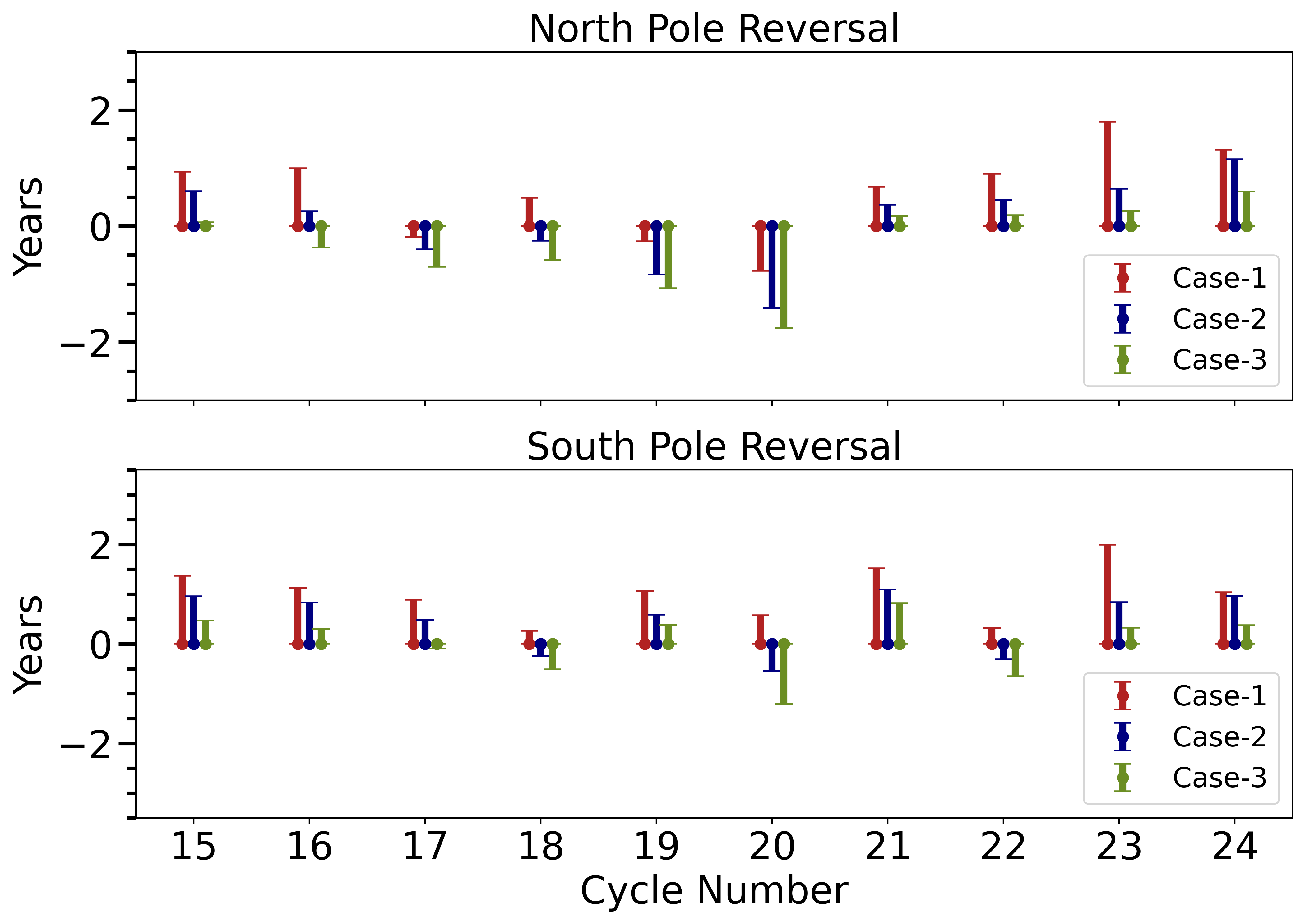}
            \caption{Comparison of the simulated pole reversal times with the observed polar field. Left panel: We compare our three cases, Case-1 (without asymmetry), Case-2 (long-term asymmetry), and Case-3 (cycle-23 asymmetry for all cycles) with observed polarity reversal from KSO data up to 1976 and WSO data after that. The deviations from observed reversal for three cases are shown using red, green, and blue vertical error bars, respectively. Right Panel: Same as left panel but for observed polar field from MWO facule data up to 1976 and WSO after 1976. }
            \label{Time_reversal}
    \end{figure*}

    We first calculate the polar field reversal time from the yearly averaged observed data for each cycle. If our simulated polar field reversal time exactly matches the observed field reversal time, we mark it as zero. If the simulated reversal time is behind the observed one, we mark it as negative, and if it is ahead of the observed time, then we mark it as positive. Figure~\ref{Time_reversal} presents this comparison. The left panel of Figure~\ref{Time_reversal} shows the comparison of polar field time reversal from KSO data \citep{Mishra2025} and the right panel shows the same with MWO data \citep{Munoz2012}. Although for both panels, the polar field data is merged with WSO data after 1976. The red error bars show the deviation of polar field reversal from observation for Case-1 without asymmetry. The blue and green error bars indicate the deviation of the simulated polar field reversal times from the observed values corresponding to simulations for Case-2 with long-term temporal variation of the morphological asymmetry factor and Case-3 with cycle 23 asymmetry, respectively.

    It can be seen from Figure~\ref{Time_reversal} that the inclusion of the morphological asymmetry factor improves the agreement between the simulated and observed pole reversal for both KSO and MWO data in most of the cycles, with a very few exceptions. The polarity reversal at the south pole becomes close to observation without asymmetry for cycle 18 in KSO data. For MWO data, the cycles 17 and 19 in the north pole reversal give better results for the case without asymmetry. For the cycles 21 -24, where polar field data is available from WSO data, the improvement in polar field reversal with asymmetry (Case-2 \& Case-3) could be more trustworthy, as observed polar field data is more reliable during this period. However, for cycle-22, our simulation shows quicker south pole reversal than observational reversal for the asymmetric cases compared to the case without asymmetry.

    When we compare our simulated polar field with KSO polar field data before 1976, the asymmetry using only cycle-23 (Case-3) gives better improvement in polar field reversal time than Case-2 with long-term asymmetry. This is also true for cycles 21-24 with WSO data. If morphological asymmetry is stronger in a cycle, then asymmetric sunspots give rise to quicker and stronger polar fields in comparison to the symmetric sunspots (See Appendix B for details). For Case-3, the asymmetry parameters (see Figure~\ref{asymmetry_factor_cycle 23} in Appendix~\ref{app:A}) introduced in the sunspots pair are stronger than in Case-2, and as a result, we expect a quicker pole reversal and close to the observational value. Although this is generally true, apparently it might look like there is some exception in Figure~\ref{Time_reversal}. For example, if we see cycle 18 for the KSO data and cycle 22 in the WSO data in the southern hemisphere (left panel of Figure~\ref{Time_reversal}), it shows that Case-2 matches better than Case-3. This is because the asymmetry parameters for Case-2 during those cycles are close to Case-3 asymmetry. The MWO data also shows similar behavior.

    \subsubsection{Peak amplitude of polar field}
\begin{figure*}[!htbp]
            \centering
            \textbf{~~~~~~~~~~~KSO data \citep{Mishra2025}~~~~~ ~~~~~~~~~~~~   ~~~~~~MWO data \citep{Munoz2012}}\par\medskip
            \includegraphics[width=0.49\linewidth]{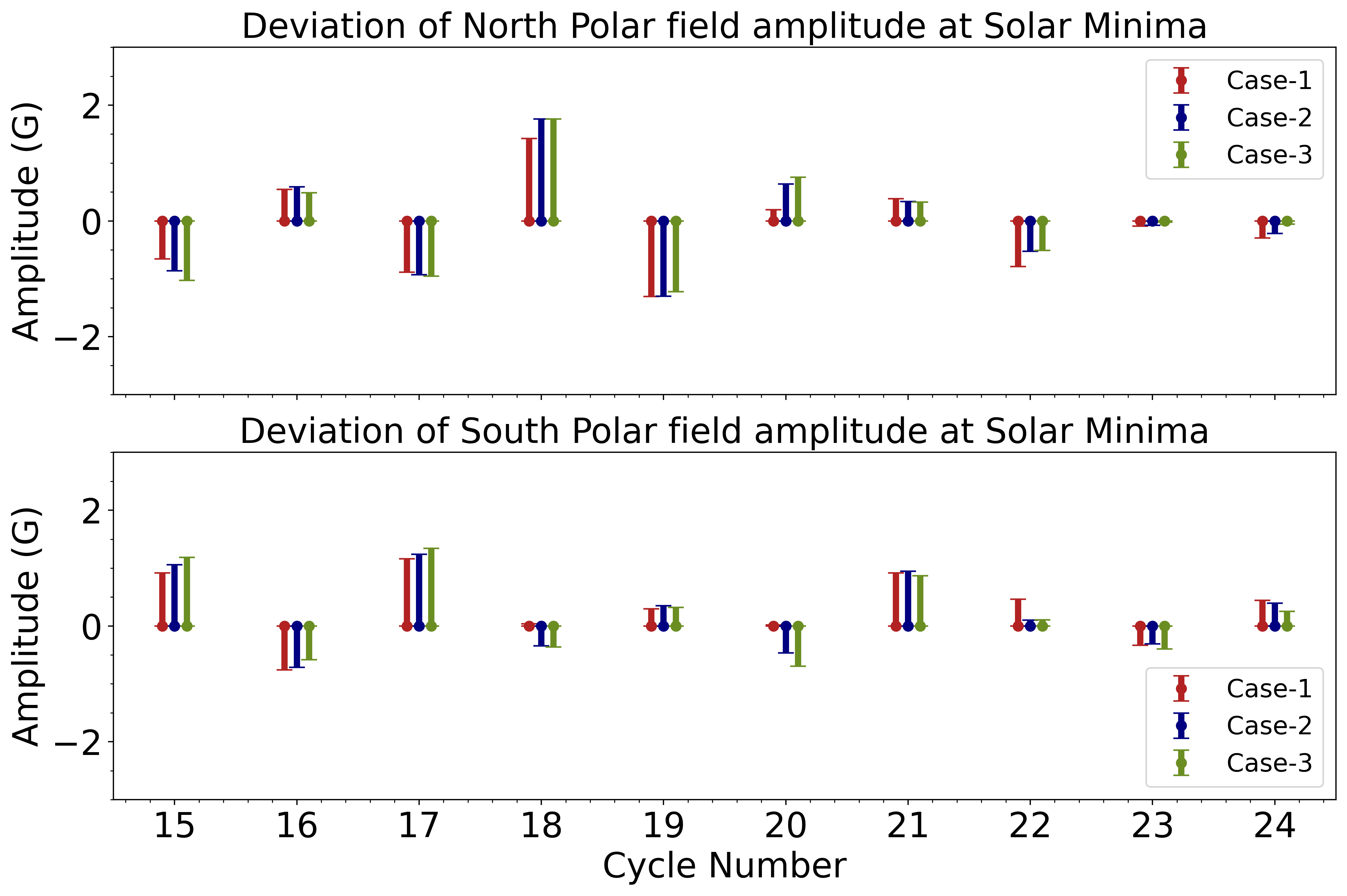}
            \includegraphics[width=0.49\linewidth]{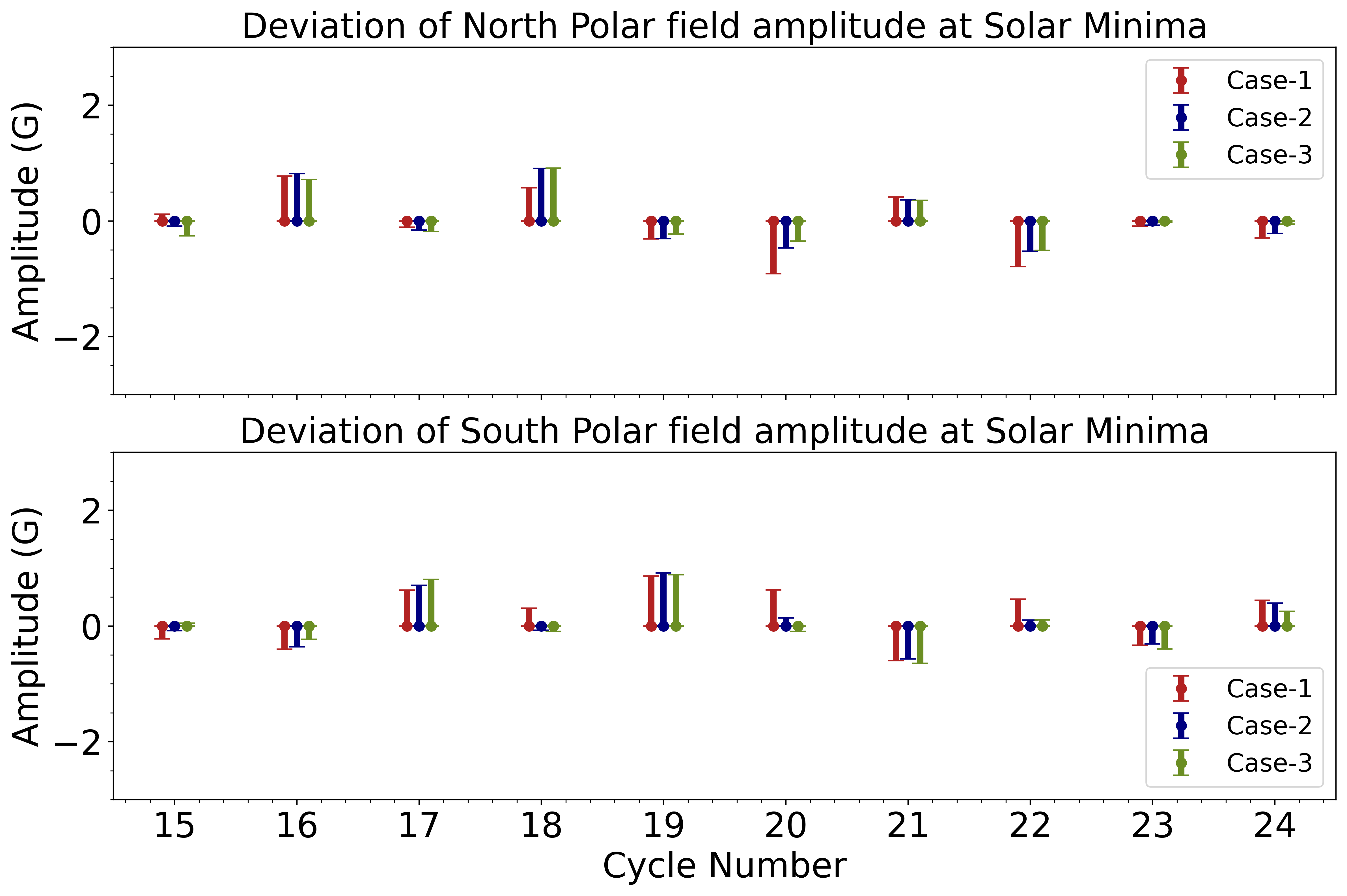}
            \caption{Comparison of the simulated peak amplitude of the polar field with the observed one for the three cases we considered in our simulations. Left Panel: The red, blue, and green error bars show the deviation of peak amplitude of simulated polar field from the observed peak amplitude from KSO data (up to 1976) and WSO data (1976-2016) for Case-1 (without asymmetry), Case-2 (Long-term asymmetry) and Case-3 (cycle 23 asymmetry), respectively. Right panel: Same as left panel, but for MWO data up to 1976 and WSO after 1976.}  
            \label{fig:peak_amplitude}
    \end{figure*}

    We also performed a comparison of the peak amplitude of the polar field during solar minima with observations. Figure~\ref{fig:peak_amplitude} shows the deviation of peak amplitude of polar fields for all three cases from observations at solar minima. More error bars indicate a higher deviation of the simulated polar field at solar minima from observational data. The blue and green error bars represent the deviation of the simulated polar field with both asymmetries (Case-2 and Case-3) considered in our study. The red error bar shows the deviation of the simulated peak amplitude of the polar field from the observed peak amplitude for the case without any asymmetry (Case-1). The left panel and right panel show the comparison of the simulation with KSO data and MWO data, respectively. It can be seen from the Figure~\ref{fig:peak_amplitude} that at the solar minima, our simulated peak amplitude of polar field for all cases (both with and without the asymmetry factor) has similar behavior. The cases with asymmetry (Case-2 \& Case-3) here do not show any dramatic difference in the peak amplitude behavior in comparison to the case without asymmetry (Case-1). Statistically, the peak amplitude of the polar field with asymmetry is a bit higher side in comparison to the no asymmetry case for most of the cycle, as the diffuse following sunspot in the asymmetry case produces a stronger polar field. A comparison with MWO data (right panel of Figure~\ref{fig:peak_amplitude}) shows a similar behavior to the KSO data. 

\section{Discussion and Conclusion}
In this study, we have demonstrated the effect of morphological asymmetry between the leading and following polarities of a Bipolar Magnetic Region (BMR) on the long-term evolution of the solar magnetic field using a modified SFT model. We introduced, for the first time, a long-term asymmetry factor in sunspots, spanning a century-long time scale (1913 to 2016), to study its impact on the evolution of the solar magnetic field on the solar surface.

 All together we performed simulations with two time-dependent asymmetries - the first one, with long-term asymmetry obtained from observations (Case-2), and the second one with time-dependent asymmetry specific to cycle-23 applied to all other cycles (Case-3). Please note that different solar cycle has different strengths and durations, and applying cycle-23 asymmetry in all other cycles is not physically consistent. We use it as a case study to understand the importance of different asymmetries on our final results. To compare our results with morphological asymmetry (Case-2 \& Case-3), we have a standard case without considering any type of asymmetry in sunspots (Case-1)  

The surface evolution of the solar magnetic field gets improved, and results become close to the observations when we consider the morphological asymmetry for both Case-2 and Case-3 (middle and lower panel of Figure~\ref{butterfly diagram}). The magnetic flux in the latitudes below $< 55^{\circ}$ is stronger for cases with asymmetry compared to the symmetric case. This improved surface evolution of magnetic flux transport is tremendously important as it provides the base for modelling the solar wind, coronal and heliospheric magnetic field \citep[e.g.,][]{Barnes2023, Huang2023, Asvestari2024}. Also, to understand the temporal evolution of coronal holes, the estimation of the middle latitude magnetic field is very crucial \citep{Watanabe2019}, which could be provided from our improved surface evolution of magnetic flux, as the magnetic flux in the low and mid latitude activity belt is close to observation for cases with asymmetry.  

Our findings also indicate that the inclusion of an asymmetry factor improves the agreement between the simulated polar field reversal time and observations in most of the cycles. Consideration of asymmetry parameters helps pole reversal close to the observed value and earlier than the case where no asymmetry is considered. We also compare the peak polar field amplitude with observations. For peak amplitude, we do not find any significant difference between our simulations with asymmetry cases and the no asymmetry case. However, for recent cycles 22-24, a very careful scrutiny would show that cycle asymmetry gives slightly better results than the no asymmetry case. This might be due to less uncertainty in many measured quantities that go into the SFT models (e.g., asymmetry, tilt angle, sunspot area measurements) for recent cycles. Overall, the inclusion of asymmetry in sunspots improves the simulation results and reproduces the surface evolution of the magnetic field and behaviour of the polar field closely to observations.

{Now the question is whether the computed 1-2 years difference in polar field reversal and a few Gauss difference in polar field strength for morphological asymmetry cases matter! It has already been established that the differences in polar field reversal time and peak amplitude of polar field have a significant effect on the peak amplitude of the next cycle \citep{Jiang2007, Munoz2012, Hazra2019, Kumar2022}, its length \citep{Solanki2002, Vaquero2008} and timing for solar maxima \citep[e.g.,][]{Bhowmik2018, Jha2024}. Also, as the SFT model alone does not predict the next cycle amplitude, we need to focus on coupled SFT \& dynamo models \citep[e.g.,][]{Bhowmik2018} or 3D model \citep{Chatterjee2026} where surface flux transport and magnetic field generation due to dynamo happen simultaneously to quantify the effect of differences in polar field on the peak-amplitude of the next cycle. However, from the observation, we can infer the effect of these differences in peak polar field on the next cycle amplitude. All the polar field proxies show a strong linear correlation between polar field and peak sunspot numbers/areas. Recently \citet{Kumar2022} revisited all available polar field data and showed strong correlations of sunspot area of next cycle with peak polar field from WSO data in northern and southern hemispheres with correlation coefficients of 0.98 and 0.76, respectively. These high correlations show that if there is an overestimate of polar field, for example, by 25\%, the predicted amplitude would be overestimated by nearly 25\%. Also, it will affect the length of the cycle as a stronger cycle tends to be shorter in length \citep{Solanki2002, Upton2018}. Therefore, overestimation or underestimation of the polar field leads to stronger (shorter) or weaker (longer) cycles.

Quantifying the effect of polar field reversal time is very crucial for predicting the timing of the next solar cycle maximum. The polar field reversal time coincides with the solar cycle maxima \citep{Choudhuri2007, Munoz2015, Mishra2025}, and hence predicting correctly when the polar field will reverse is eventually a task of predicting the timing of the solar maxima. There are many groups \citep[e.g.,][]{Upton2018, Jha2024} who put efforts into predicting polar field reversal time using SFT models. Predicting polar field reversal using SFT is a bit challenging as it involves many convective transport parameters (meridional circulation, diffusion, convective transport by granular and supergranular cells) that are not well constrained. It also depends on how one models the bipolar magnetic regions, whether the asymmetry is considered or not, as we have shown in our study. Hence, it is extremely important to correctly compute the polar field reversal time and peak amplitude, as polar field time reversal and amplitude decide the next cycle amplitude and timing \citep{Choudhuri2007, Jiang2007, Jiang2013, Cameron2016, Hazra2019}. Any discrepancy in the computed polar field reversal time and its amplitude will lead to a problematic prediction for the next solar cycle.}


In all of our calculations, if we compare which type of asymmetry gives us a better polar field evolution in terms of pole reversal, then it becomes subjective on which type of observational data set we are using to compare our results with. For example, if we compare the polar field reversal time (Figure~\ref{Time_reversal}) with KSO polar field data, the cycle 23 asymmetry (Case-3) gives a relatively better result compared to the long-term asymmetry (Case-2). On the other hand, if we compare our simulation results with MWO data, it is not easy to draw a conclusion. {For the north pole, we found that difference in the pole reversal time between these two observational datasets is about 1-2 years for most of the cycles. However, for the south pole, the difference is small in most cases, less than or equal to 1 year. The observational uncertainly in the polar field measurement is a bottleneck to correctly quantify the effect of morphological asymmetry in the polar field and its reversal.} In hindsight, long-term asymmetry should be more physical, as different solar cycles should have different asymmetries, as the strengths of the cycles are different. Also, for Case-3, we have incorporated the asymmetry for both hemispheres differently, but for long-term asymmetry in Case-2, we use the same asymmetry profile as shown in Figure~\ref{fig:long_term_fspot} for both northern and southern hemispheres. Nevertheless, both the asymmetry cases give better results (close to observation) compared to the no asymmetry case. When we compare our simulations with WSO data after 1976, cycle-23 asymmetry gives better results too. In all of our simulations, we see differences in the polar field reversal time and peak amplitude of the polar field in the northern and southern hemispheres. This suggests that accurately accounting for the hemispheric variation of the asymmetry factor might be crucial for improving the predictive capability of surface flux transport models. We have also not considered the latitudinal dependence of sunspot asymmetry in our calculation, which might have some effect on the overall polar field evolution. Future study should include these. 

{Although our study shows that incorporating asymmetry correctly predicts polar field reversal and peak amplitude close to observations, one might wonder how our results depends on the various parameters that we assume in our model, in particular, the decay time ($\tau$) for radial diffusion, the flow profile, the super-granular diffusion coefficient ($\eta_H$) and tilt angles. All might influence the outcome of our study. Below, we mention the dependence of our results on these parameters.  

\subsection{Effect of decay time ($\tau$)}
We carried out a set of test cases to examine the effect of the decay time ($\tau$) on our results. In our simulation so far, we have assumed ($\tau$) to be $10.2$ years. To compare how our results would change for different ($\tau$), we have considered $\tau = 4.5$ years and $\tau = 20$ years, keeping all other parameters unchanged. For $\tau = 4.5$ years, the value of $B_T$ becomes higher, while for $\tau = 20$ years, $B_T$ becomes lower compared to the normalized value of $B_T$ that we obtained by normalizing the polar field strength of cycle 21. This is expected because a smaller $\tau$ represents no magnetic memory of the last cycle, causing the flux to decay more rapidly. In contrast, a larger $\tau$ allows the flux to decay slowly, so the total flux remains higher for a longer time, resulting in a higher polar field. The decay time also has a clear impact on the polar field reversal. A short decay time (weak memory) leads to an earlier reversal, while a long decay time delays the reversal, causing the poles to switch after the actual reversal time.

A short decay time removes the diffused field more quickly than a long decay time, which affects the peak amplitude of the polar field. Although the changes in the peak amplitude are not very large, the inclusion of asymmetry with a longer decay time provides a closer match to the observed polar field compared to the no-asymmetry case for most cycles. A short decay time also reduces the influence of asymmetry, making the polar field of the asymmetry and non-asymmetry cases appear quite similar.

\subsection{Effect of meridional flow}
To understand the effect of the meridional flow, we also performed a set of test cases by changing only its amplitude. Our simulation so far used a maximum meridional flow speed of $10.06~\text{ms}^{-1}$. We considered two additional cases with amplitudes of $9.2~\text{ms}^{-1}$ and $12~\text{ms}^{-1}$, keeping the shape of the flow profile unchanged. Compared to the normalized value of $B_T$ using Cycle~21, its value becomes higher for the $9.2~\text{ms}^{-1}$ case and lower for the $12~\text{ms}^{-1}$ case.
This behavior is quite expected because a faster meridional flow transports the following polarity flux to the poles more quickly, producing a stronger polar field. A slower flow delays this transport and results in a weaker polar field. These differences strongly affect the timing of the polar field reversal. With a fast flow, the higher polar field reaches the poles earlier, causing an early reversal relative to observations but a slow flow delays the arrival of flux at the poles, leading to a delayed reversal. This is in accordance with the earlier studies \citep[e.g.,][]{Whitbread_2017}

To examine the effect of the velocity profile structure, we performed two additional cases in which the meridional flow amplitude was kept at $10.06~\text{m s}^{-1}$, but the latitude of the peak flow was changed. In our model, the peak of meridional flow occurs at $33.2^\circ$, while for the two test cases that we consider, it peaks at $23.9^\circ$ and $43.5^\circ$ (see Figure~\ref{fig:mf_lat}). When the peak is at $23.9^\circ$, the meridional flow transports much of the following polarity together with the leading polarity, resulting in strong cancellation between the two. To compensate for this cancellation and match the observed polar field, a much larger value of $B_T$ is required. In contrast, when the peak is located at $43.5^\circ$, very little leading polarity is carried poleward, so the cancellation is weaker and the required value of $B_T$ is slightly lower.

\begin{figure}[!htbp]
            \centering
            \includegraphics[width=0.9\linewidth]{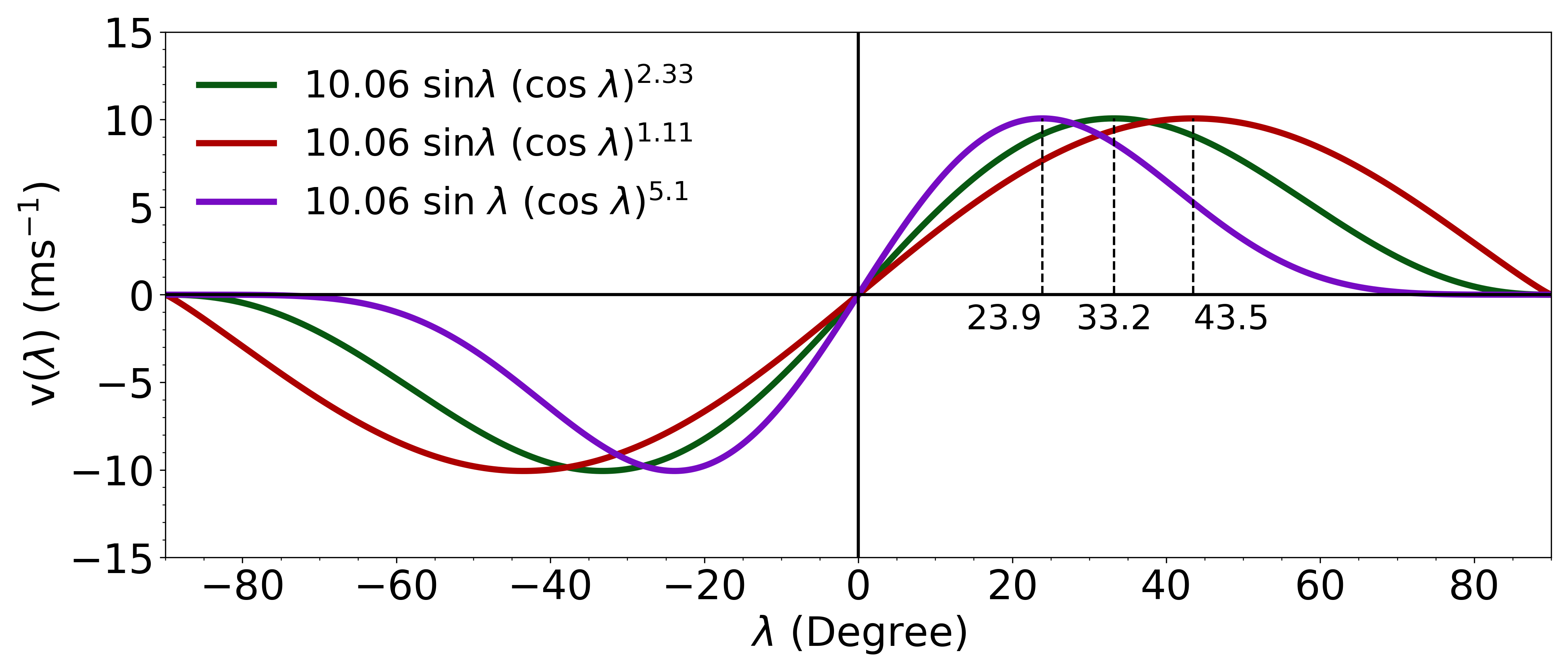}
            \caption{Meridional flow profiles with a maximum speed of $10.06~\text{ms}^{-1}$. The green, red, and purple lines show three different profiles of the meridional flow. The flow reaches its highest value at about $\pm 33.2^{\circ}$, $\pm 43.5^{\circ}$, and $\pm 23.9^{\circ}$ latitude for each profile, respectively.}
            \label{fig:mf_lat}
\end{figure}
    
The different flow profiles also influence the timing of the polar field reversal. With the peak at $43.5^\circ$, the reduced cancellation allows more following polarity to reach the poles, causing a slightly earlier reversal. For the peak at $23.9^\circ$, strong cancellation leaves less following polarity to be transported poleward, leading to a delayed reversal. Although the peak amplitudes do not change significantly between the cases, the differences between the asymmetry and the no-asymmetry cases are slightly smaller when the meridional flow peaks at $43.5^\circ$ compared to the $23.9^\circ$ case.

\subsection{Effect of diffusion coefficient ($\eta_H$)}
We performed another set of simulation using $\eta_H = 455.7~\text{km}^2\text{s}^{-1}$, where the result presented in our study uses $\eta_H = 250~\text{km}^2\text{s}^{-1}$. Due to the higher diffusion coefficient, we increased $B_T$ to match the polar field strength of Cycle~21. We found that increased diffusion significantly reduced the impact of asymmetry, resulting in solutions that closely resemble the symmetric case. The poles also reverse slightly earlier in this high-diffusion scenario. In addition, the differences in peak amplitudes for the asymmetry and non-asymmetry cases become insignificant.  

An estimation of the proper tilt angle is very crucial to quantify the effect of morphological asymmetry in the polar field. Unfortunately, we do not have proper estimates of the tilt angle of sunspots, especially for earlier solar cycles. An improper tilt angle combined with asymmetric sunspots gives rise to a different polar field (see Appendix~\ref{app:B} for details). As shown in Table~\ref{Tilt_table}, up to Cycle 20, there is no specific tilt angle data available separately for the northern and southern hemispheres; only averaged data is available. This lack of precise historical data may introduce uncertainties into SFT simulations, affecting the accuracy of the modeled polar field evolution.}
    
Despite these limitations, the inclusion of the morphological asymmetry and its long-term modulation over the solar cycle is crucial for the reconstruction of the solar magnetic field using SFT models. This improves the polar field reversal time and spatio-temporal variation of surface magnetic field for all cycles in the last century, especially for the last three solar cycles (Cycles 21, 22, and 23). Our results suggest that the accuracy of BMR-based flux transport models can be further enhanced by incorporating more detailed hemisphere-specific asymmetry over all cycles and improved tilt angle measurements. Future work should focus on refining these parameters to achieve a more comprehensive understanding of the factors governing polar field evolution and reversal.

\appendix
    \section{Asymmetry factor based on Cycle 23}
    \label{app:A}
    In our study, we examined a specific case (Case-3) with cycle-23 asymmetry applied to all cycles on the overall evolution of the surface magnetic field. The cycle-23 asymmetry factor was determined based on the monthly average of the following and leading polarity of sunspots during cycle 23. However, due to the unavailability of asymmetry factor data for individual solar cycle, we relied on the variation of the asymmetry factor observed during Cycle 23 (\cite{Tlatov2014}). 
    
\begin{figure}[!htbp]
            \centering
            \includegraphics[width=1\linewidth]{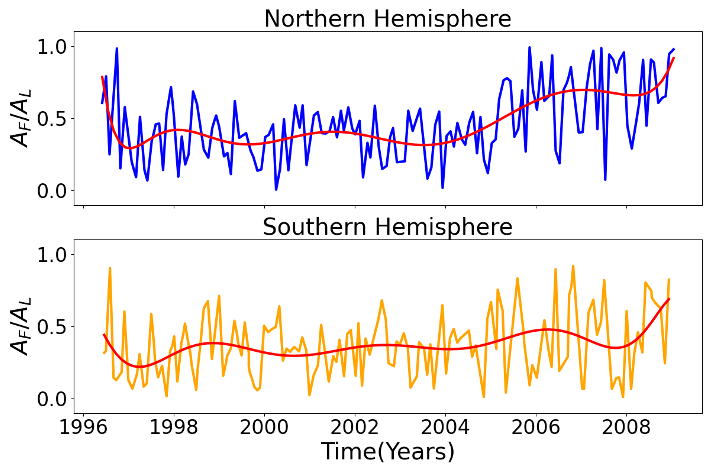}
            \caption{Temporal variation of the asymmetry factor during Solar Cycle 23. The upper panel displays the temporal variation of the asymmetry factor for the Northern Hemisphere, while the lower panel presents the corresponding values for the Southern Hemisphere. The red curves in both panels represent polynomial fits to the temporal variation of the asymmetry factor.}
            \label{asymmetry_factor_cycle 23}
    \end{figure}

\begin{figure*}[!htbp]
                \centering
                \includegraphics[width=0.9\linewidth]{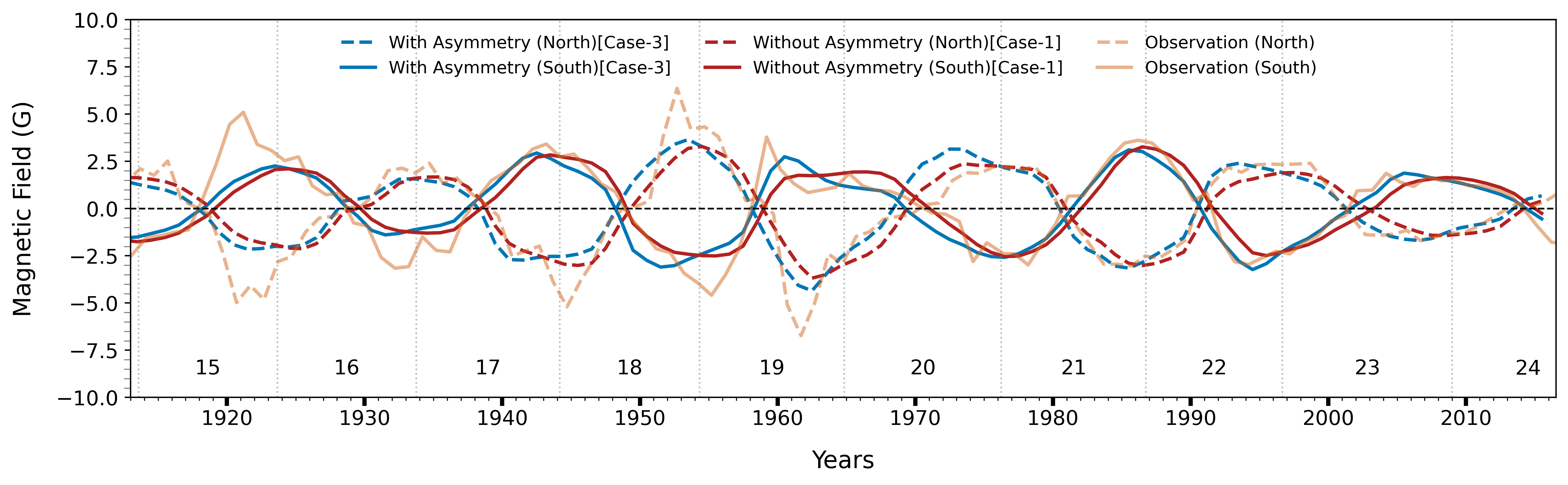}
                \caption{Comparison of the polar field over 100 years from the simulation, with and without the temporally varying asymmetry factor based on Cycle 23, against observational MWO (1913 - 1976) and WSO (1976 - 2016)} polar field data. The solid line represents the south pole, while the dotted line represents the north pole
                \label{polar field cycle 23}
    \end{figure*}

    Here, in Figure \ref{asymmetry_factor_cycle 23}, we fitted the asymmetry factor for both the southern and northern hemispheres. We used this hemisphere-specific fitted value in every cycle in our simulation. The century-long simulation of the polar field has been shown in Figure~\ref{polar field cycle 23}.
    The errors in the pole reversal time and on the amplitude at the solar minima compared to the observations have been discussed in the previous section~\ref{sec:polarfield}.


    \section{Effect of a {single bipolar sunspot} on the polar field with and without inclusion of the asymmetry factor} \label{app:B}
\begin{figure}[!htbp]
        \centering
        \begin{subfigure}{1\linewidth}
            \captionsetup{position=top} 
            \caption{North polar magnetic field}
            \centering
            \includegraphics[width=1\linewidth]{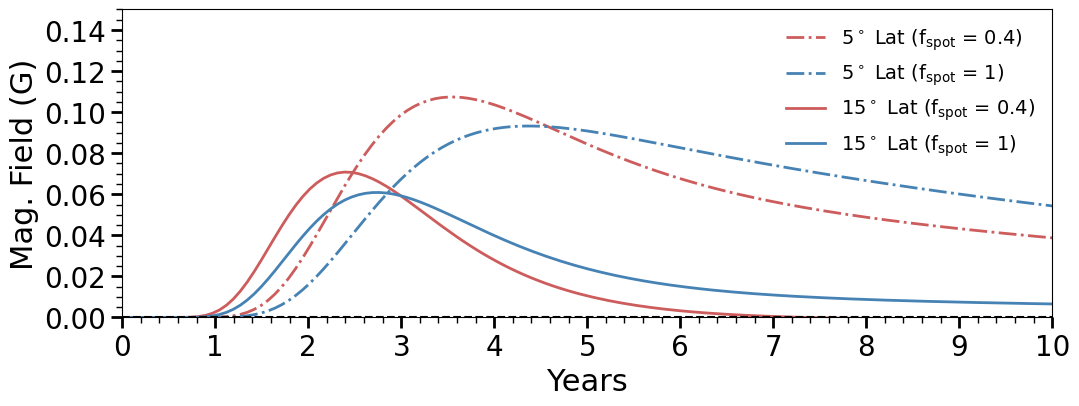}
            \label{fig:single_pole_3tilt_north}
        \end{subfigure}
        
        
        \begin{subfigure}{1\linewidth}
            \captionsetup{position=top}
            \caption{South polar magnetic field}
            \centering
            \includegraphics[width=1\linewidth]{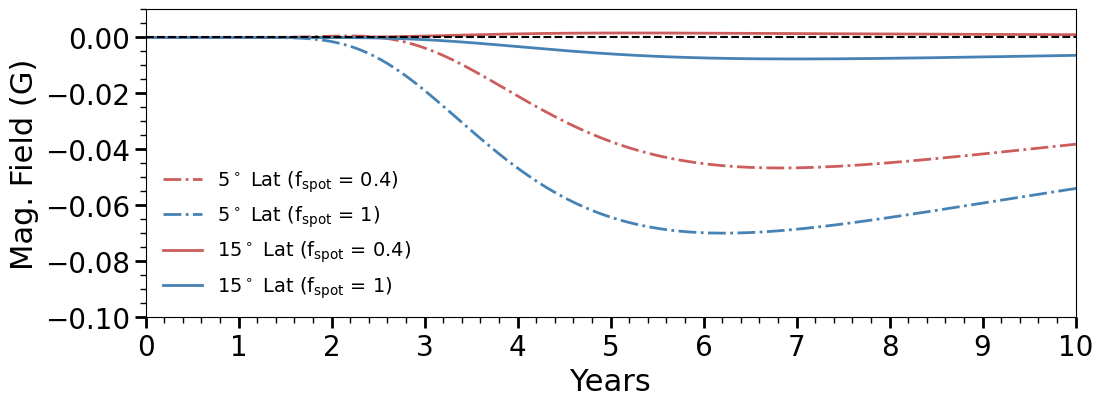}
            \label{fig:single_pole_3tilt_south}
        \end{subfigure}
        
        \caption{Temporal evolution of the Sun’s polar magnetic field following the emergence of a single bipolar sunspot region at two different latitudes: $5^\circ$ and $15^\circ$, with a tilt angle of $3^\circ$ and a total sunspot area of $2000\ \mu Hm$. Two asymmetry factors are compared: $\text{f}_\text{spot} = 1$ (symmetric) and $\text{f}_\text{spot} = 0.4$ (asymmetric). Panel (a) shows the comparison of the north polar field, while panel (b) shows the comparison of the south polar field.}
        \label{fig:single_pole_3tilt}
    \end{figure}

    Here we study how the Sun’s polar magnetic field evolves due to a single sunspot emerging at different latitudes at $5^\circ$ and $15^\circ$, for both symmetric and asymmetric cases. We consider a sunspot with an area of $2000\ \mu Hm$ and a tilt angle of $3^\circ$ (see Figure~\ref{fig:single_pole_3tilt}). As shown in Figure~\ref{fig:single_pole_3tilt_north}, the north polar field initially becomes stronger in the asymmetric case. This happens because the following polarity is more spread out and diffuse, while the leading polarity remains compact. As a result, less cancellation occurs between the two polarities at the beginning of the evolution, allowing more following magnetic flux to be transported to the pole compared to the symmetric case. Initially, the north polar field strengthens more rapidly in the asymmetric case due to reduced cancellation and faster poleward transport of the following polarity. However, as the leading polarity begins to reach the north pole, it encounters and start canceling the flux of the following polarity. Because a portion of the following polarity has diffused into the opposite hemisphere due to cross-equatorial diffusion, less flux remains in the north, resulting in a sharper decline in the polar field compared to the symmetric case.
    
    At low latitudes (e.g., $5^\circ$), the asymmetric case shows stronger cross-equatorial diffusion (see Figure ~\ref{fig:single_pole_3tilt_south}). This is because, near the equator, the meridional flow has a weaker effect and with a low tilt angle, its influence is nearly the same for the two polarities. Also, the following polarity is more spread out due to its larger size, allowing more of it to diffuse across the equator. This results in a weaker polar field in the south pole compared to the symmetric case. If we look at the build-up of the south polar field carefully, there will be a major contribution from the leading polarity as well as a little contribution from the following polarity from the northern hemisphere due to cross-equatorial diffusion. In case of asymmetry, cross-equatorial diffusion of the following polarity becomes more, in turn it cancels more of the leading polarity flux, resulting in less polar field than the symmetric case. 
    
    At higher latitudes ($15^\circ$), the situation remains the same. Here, the meridional flow is more effective and carries much of the following polarity. Therefore, the cross-equatorial diffusion of the following polarity is reduced, leading to less cancellation between the polarities in the southern hemisphere for the asymmetric case compared to low-latitude emergence. As a result the difference between the polar field for symmetric and asymmetric case gets very close to each other.

\begin{figure}[!htbp]
    \centering

    \begin{subfigure}{1\linewidth}
        \captionsetup{position=top} 
        \caption{North polar magnetic field}
        \centering
        \includegraphics[width=1\linewidth]{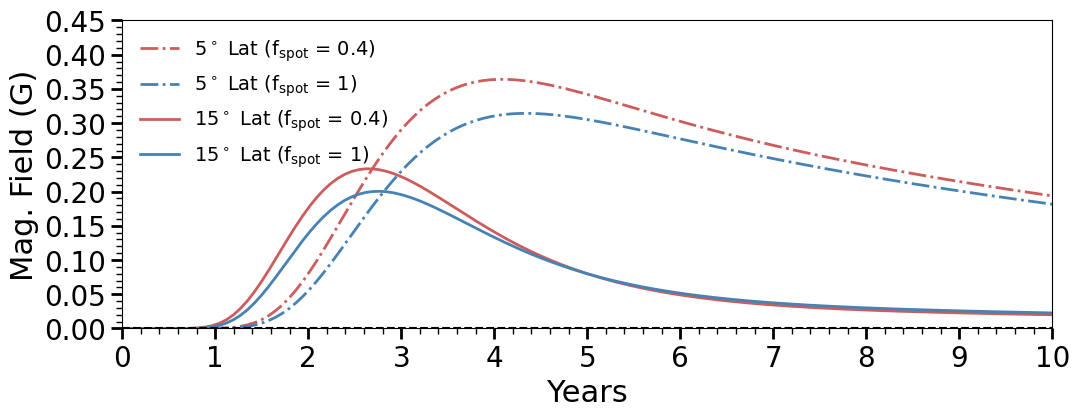}
        \label{fig:single_pole_10tilt_north}
    \end{subfigure}
    
    
    \begin{subfigure}{1\linewidth}
        \captionsetup{position=top} 
        \caption{South polar magnetic field}
        \centering
        \includegraphics[width=1\linewidth]{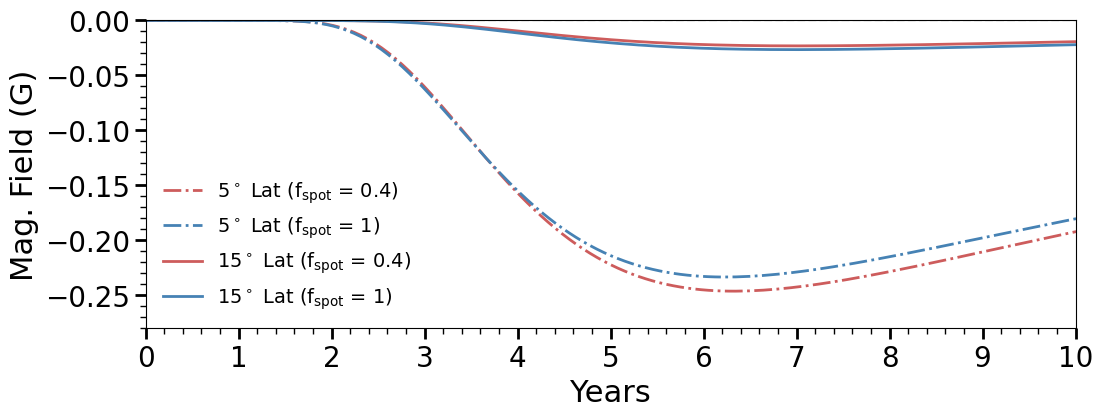}
        \label{fig:single_pole_10tilt_south}
    \end{subfigure}
    
    \caption{Same as Figure~\ref{fig:single_pole_3tilt} with tilt angle $10^\circ$. }
    \label{fig:single_pole_10tilt}
    \end{figure}

    We also performed another case study by increasing the tilt angle to $10^\circ$, which is almost triple than the previous case, following the same setup as in the previous analysis. For the north and south polar field, we observe a similar trend as before (see Figure~\ref{fig:single_pole_10tilt}). However, the difference in the south polar field (see Figure~\ref{fig:single_pole_10tilt_south}) gets less between the symmetric and the asymmetric case. This is because, at higher tilt angles, the meridional flow affects the leading and following polarities differently, reduce cross equatorial diffusion by taking most of the following polarity to the north pole. From Figure~\ref{fig:single_pole_10tilt_south} it can be seen that for $15^\circ$ emergence latitude the cross-equatorial diffusion is a little bit less for asymmetric case compared to the symmetric one because hight tilt angle and high meridional flow at upper latitude takes most of the diffused polarity than the compact one.
    This clearly shows the effect of tilt angle along with asymmetry factor is very important for the polar field development.
{
\section{Effect of isolated single-sunspot BMRs on the polar field} \label{app:C} 
In this appendix, we explore a very intriguing question whether some explicit fraction of isolated single-sunspot BMR present in total number of BMRs can have similar effect as the asymmetry of Sunspots. In our BMR dataset, we assume that all sunspots belong to bipolar magnetic regions (BMRs) and the peak magnetic strength of each polarity must be greater than the threshold value $B_T$. This ensures that both the leading and following polarities are present and clearly detectable. As a result, isolated single-polarity (monopolar) sunspots do not naturally appear in our simulations. If we introduce a purely single-polarity sunspot (a true monopole) into the simulation (by forcefully making the area of one polarity to be zero), magnetic flux is no longer balanced. The excess flux is transported toward the poles, where it accumulates and artificially modifies the polar magnetic field. Although strict flux balance is not mathematically required in a 2D surface flux transport simulation, in practice we find that flux imbalance significantly distorts the evolution of the polar field. Therefore, to obtain a physically reasonable and stable polar field, it is necessary to assume bipolar magnetic regions.

To investigate the possible impact of a single sunspot in a controlled way, we do not insert true monopoles. Instead, we consider an extreme asymmetry case. In this approach, we select an explicit fraction of sunspots and set their asymmetry factor to values close to zero. This makes the following polarity extremely weak, such that its peak field becomes very small or effectively undetectable (close to or below $B_T$). In observational terms, such regions would appear almost like single sunspots, even though a very weak opposite polarity still technically exists in the model. Using this assumption, we performed four case studies. We randomly selected $5\%$, $10\%$, $20\%$, and $30\%$ of the total BMRs incorporated in our model and assigned them with this extreme asymmetry.
Our results show the following behavior:
\begin{itemize}
    \item For $5\%$ and $10\%$, the evolution of the polar field shows almost no deviation from Case-1 (no asymmetry case with asymmetry factor $f = 1$).
    \item At $20\%$, the solution begins to deviate from Case-1 and gradually shifts toward the behavior seen in Case-2 (asymmetry case with long term asymmetry ).
    \item At $30\%$, the results become closer to the Case-2.
\end{itemize}
This approach allows us to study the sensitivity of the polar field to strong polarity imbalance while still maintaining numerical stability in the model.

\begin{figure*}[htbp]
    \centering
    \includegraphics[width=0.9\textwidth]{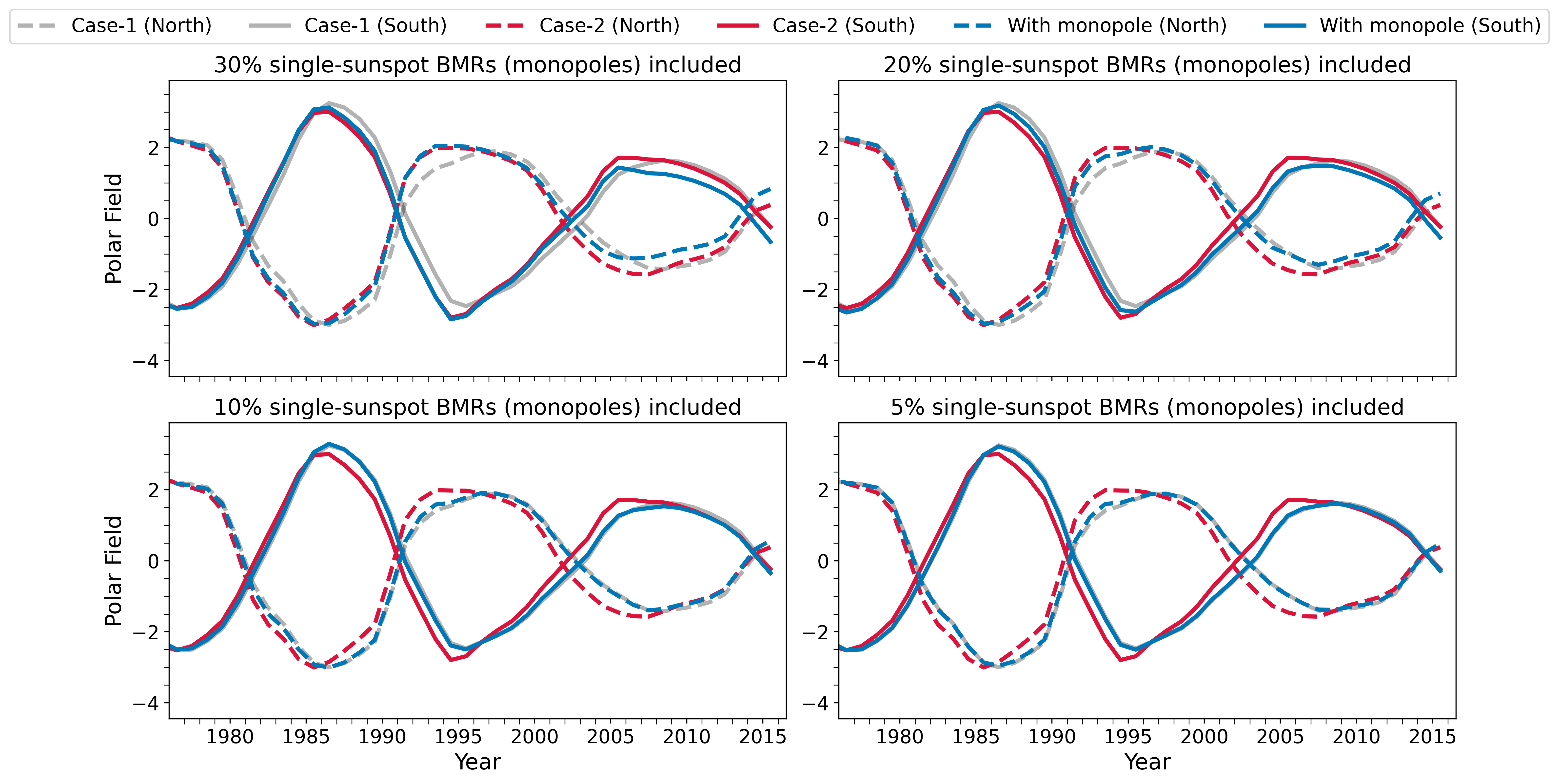}
    \caption{Importance of single-sunspot BMRs on build up of polar field for cycles 21, 22, 23. Four different cases with an explicit fraction of 30\%, 20\%, 10\% and 5\% are shown in top left, top right, bottom left and bottom right panel respectively. In each panel, the blue line represents the polar field with explicit fraction of monopole, the grey line corresponds to Case-1 with no asymmetry (of our study), and the red line corresponds to Case-2 with long-term asymmetry.}
    \label{fig:myfigure}
\end{figure*}

For a small explicit fraction of single sunspots in BMR dataset (e.g., $5\%$ and $10\%$), the results are almost identical to Case-1. As we increase the single-sunspot percentage (e.g., $30\%$), the solution becomes closer to Case-2. However, for the Cycle 23, the $30\%$ case does not exactly match Case-2. This is because different tilt angles were used for symmetric and asymmetric cases in our main study for the Cycle 23 \& Cycle 24, and in the present experiment, the tilt angles were kept the same as Case-1 to see the effect of some fraction of single-sunspot regions in total bipolar magnetic regions. }

\begin{acknowledgements}
The authors would like to thank an anonymous referee for their thoughtful comments. SP and GH acknowledge funding from IIT Kanpur initiation grant IITK/PHY/2022386. 
\end{acknowledgements}

\bibliographystyle{aasjournal}
\bibliography{bibliogrphy}

\begin{thebibliography}{}
\expandafter\ifx\csname natexlab\endcsname\relax\def\natexlab#1{#1}\fi

\bibitem[{{Asvestari} {et~al.}(2024){Asvestari}, {Temmer}, {Caplan}, {Linker},
  {Heinemann}, {Pinto}, {Henney}, {Arge}, {Owens}, {Madjarska}, {Pomoell},
  {Hofmeister}, {Scolini}, \& {Samara}}]{Asvestari2024}
{Asvestari}, E., {Temmer}, M., {Caplan}, R.~M., {et~al.} 2024, \apj, 971, 45

\bibitem[{{Barnes} {et~al.}(2023){Barnes}, {DeRosa}, {Jones}, {Arge}, {Henney},
  \& {Cheung}}]{Barnes2023}
{Barnes}, G., {DeRosa}, M.~L., {Jones}, S.~I., {et~al.} 2023, \apj, 946, 105

\bibitem[{{Baumann, I.} {et~al.}(2004){Baumann, I.}, {Schmitt, D.},
  {Schüssler, M.}, \& {Solanki, S. K.}}]{Baumann_2004}
{Baumann, I.}, {Schmitt, D.}, {Schüssler, M.}, \& {Solanki, S. K.} 2004, A\&A,
  426, 1075

\bibitem[{{Bhowmik} \& {Nandy}(2018)}]{Bhowmik2018}
{Bhowmik}, P., \& {Nandy}, D. 2018, Nature Communications, 9, 5209

\bibitem[{Cameron {et~al.}(2010)Cameron, Jiang, Schmitt, \&
  Schüssler}]{Cameron_2010}
Cameron, R.~H., Jiang, J., Schmitt, D., \& Schüssler, M. 2010, The
  Astrophysical Journal, 719, 264

\bibitem[{{Cameron} {et~al.}(2016){Cameron}, {Jiang}, \&
  {Sch{\"u}ssler}}]{Cameron2016}
{Cameron}, R.~H., {Jiang}, J., \& {Sch{\"u}ssler}, M. 2016, \apjl, 823, L22

\bibitem[{{Chatterjee} \& {Hazra}(2026)}]{Chatterjee2026}
{Chatterjee}, S., \& {Hazra}, G. 2026, \apjl, 997, L27

\bibitem[{{Choudhuri} {et~al.}(2007){Choudhuri}, {Chatterjee}, \&
  {Jiang}}]{Choudhuri2007}
{Choudhuri}, A.~R., {Chatterjee}, P., \& {Jiang}, J. 2007, \prl, 98, 131103

\bibitem[{Fan(2009)}]{Fan2009}
Fan, Y. 2009, Living Reviews in Solar Physics, 6, 4

\bibitem[{Fisher {et~al.}(2000)Fisher, Fan, Longcope, Linton, \&
  Pevtsov}]{Fisher2000}
Fisher, G., Fan, Y., Longcope, D., Linton, M., \& Pevtsov, A. 2000, Solar
  Physics, 192, 119

\bibitem[{{Fligge} \& {Solanki}(1997)}]{Fligge1997}
{Fligge}, M., \& {Solanki}, S.~K. 1997, \solphys, 173, 427

\bibitem[{{Grotrian} \& {K{\"u}nzel}(1950)}]{Grotrian1950}
{Grotrian}, W., \& {K{\"u}nzel}, H. 1950, \zap, 28, 28

\bibitem[{Hathaway {et~al.}(2003)Hathaway, Nandy, Wilson, \&
  Reichmann}]{Hathaway_2003}
Hathaway, D.~H., Nandy, D., Wilson, R.~M., \& Reichmann, E.~J. 2003, The
  Astrophysical Journal, 589, 665

\bibitem[{{Hathaway} {et~al.}(2002){Hathaway}, {Wilson}, \&
  {Reichmann}}]{Hathaway2002}
{Hathaway}, D.~H., {Wilson}, R.~M., \& {Reichmann}, E.~J. 2002, \solphys, 211,
  357

\bibitem[{{Hazra}(2021)}]{Hazra2021}
{Hazra}, G. 2021, Journal of Astrophysics and Astronomy, 42, 22

\bibitem[{{Hazra} \& {Choudhuri}(2019)}]{Hazra2019}
{Hazra}, G., \& {Choudhuri}, A.~R. 2019, \apj, 880, 113

\bibitem[{{Hofer, B.} {et~al.}(2024){Hofer, B.}, {Krivova, N. A.}, {Cameron,
  R.}, {Solanki, S. K.}, \& {Jiang, J.}}]{Hofer_2024}
{Hofer, B.}, {Krivova, N. A.}, {Cameron, R.}, {Solanki, S. K.}, \& {Jiang, J.}
  2024, A\&A, 683, A48

\bibitem[{{Huang} {et~al.}(2023){Huang}, {T{\'o}th}, {Sachdeva}, {Zhao}, {van
  der Holst}, {Sokolov}, {Manchester}, \& {Gombosi}}]{Huang2023}
{Huang}, Z., {T{\'o}th}, G., {Sachdeva}, N., {et~al.} 2023, \apjl, 946, L47

\bibitem[{Iijima {et~al.}(2019)Iijima, Hotta, \& Imada}]{Iijima_2019}
Iijima, H., Hotta, H., \& Imada, S. 2019, The Astrophysical Journal, 883, 24

\bibitem[{{Jha} \& {Upton}(2024)}]{Jha2024}
{Jha}, B.~K., \& {Upton}, L.~A. 2024, \apjl, 962, L15

\bibitem[{Jiang {et~al.}(2009)Jiang, Cameron, Schmitt, \&
  Schüssler}]{Jiang_2009}
Jiang, J., Cameron, R., Schmitt, D., \& Schüssler, M. 2009, The Astrophysical
  Journal, 709, 301–307

\bibitem[{Jiang {et~al.}(2013)Jiang, Cameron, Schmitt, \&
  Sch{\"u}ssler}]{Jiang2013}
Jiang, J., Cameron, R.~H., Schmitt, D., \& Sch{\"u}ssler, M. 2013, Space
  Science Reviews, 176, 289

\bibitem[{Jiang {et~al.}(2014{\natexlab{a}})Jiang, Cameron, \&
  Schüssler}]{Jiang_2014a}
Jiang, J., Cameron, R.~H., \& Schüssler, M. 2014{\natexlab{a}}, The
  Astrophysical Journal, 791, 5

\bibitem[{{Jiang} {et~al.}(2007){Jiang}, {Chatterjee}, \&
  {Choudhuri}}]{Jiang2007}
{Jiang}, J., {Chatterjee}, P., \& {Choudhuri}, A.~R. 2007, \mnras, 381, 1527

\bibitem[{Jiang {et~al.}(2014{\natexlab{b}})Jiang, Hathaway, Cameron,
  {et~al.}}]{Jiang2014b}
Jiang, J., Hathaway, D.~H., Cameron, R.~H., {et~al.} 2014{\natexlab{b}}, Space
  Science Reviews, 186, 491

\bibitem[{{Jiang, J.} {et~al.}(2011){Jiang, J.}, {Cameron, R. H.}, {Schmitt,
  D.}, \& {Schüssler, M.}}]{Jiang_2011}
{Jiang, J.}, {Cameron, R. H.}, {Schmitt, D.}, \& {Schüssler, M.} 2011, A\&A,
  528, A82

\bibitem[{{Jiao, Qirong} {et~al.}(2021){Jiao, Qirong}, {Jiang, Jie}, \& {Wang,
  Zi-Fan}}]{Jiao_2021}
{Jiao, Qirong}, {Jiang, Jie}, \& {Wang, Zi-Fan}. 2021, A\&A, 653, A27

\bibitem[{{Kumar} {et~al.}(2022){Kumar}, {Biswas}, \& {Karak}}]{Kumar2022}
{Kumar}, P., {Biswas}, A., \& {Karak}, B.~B. 2022, \mnras, 513, L112

\bibitem[{{Leighton}(1964)}]{Leighton64}
{Leighton}, R.~B. 1964, \apj, 140, 1547

\bibitem[{{Mandal} {et~al.}(2017){Mandal}, {Hegde}, {Samanta}, {Hazra},
  {Banerjee}, \& {Ravindra}}]{Mandal2017}
{Mandal}, S., {Hegde}, M., {Samanta}, T., {et~al.} 2017, \aap, 601, A106

\bibitem[{{Mandal} {et~al.}(2020){Mandal}, {Krivova}, {Solanki}, {Sinha}, \&
  {Banerjee}}]{Mandal2020}
{Mandal}, S., {Krivova}, N.~A., {Solanki}, S.~K., {Sinha}, N., \& {Banerjee},
  D. 2020, \aap, 640, A78

\bibitem[{{Mishra} {et~al.}(2025){Mishra}, {Jha}, {Chatzistergos}, {Ermolli},
  {Banerjee}, {Upton}, \& {Khan}}]{Mishra2025}
{Mishra}, D.~K., {Jha}, B.~K., {Chatzistergos}, T., {et~al.} 2025, \apj, 982,
  78

\bibitem[{{Mu{\~n}oz-Jaramillo} {et~al.}(2015){Mu{\~n}oz-Jaramillo},
  {Senkpeil}, {Windmueller}, {Amouzou}, {Longcope}, {Tlatov}, {Nagovitsyn},
  {Pevtsov}, {Chapman}, {Cookson}, {Yeates}, {Watson}, {Balmaceda}, {DeLuca},
  \& {Martens}}]{Munoz2015}
{Mu{\~n}oz-Jaramillo}, A., {Senkpeil}, R.~R., {Windmueller}, J.~C., {et~al.}
  2015, \apj, 800, 48

\bibitem[{Murak\"ozy {et~al.}(2014)Murak\"ozy, Baranyi, \&
  Ludm\'any}]{Murakozy2014}
Murak\"ozy, J., Baranyi, T., \& Ludm\'any, A. 2014, Solar Physics, 289, 563

\bibitem[{Muñoz-Jaramillo {et~al.}(2012)Muñoz-Jaramillo, Sheeley, Zhang, \&
  DeLuca}]{Munoz2012}
Muñoz-Jaramillo, A., Sheeley, N.~R., Zhang, J., \& DeLuca, E.~E. 2012, The
  Astrophysical Journal, 753, 146

\bibitem[{{Pal} \& {Nandy}(2025)}]{PN2025}
{Pal}, S., \& {Nandy}, D. 2025, \aap, 700, L15

\bibitem[{{Scherrer} {et~al.}(2012){Scherrer}, {Schou}, {Bush}, {Kosovichev},
  {Bogart}, {Hoeksema}, {Liu}, {Duvall}, {Zhao}, {Title}, {Schrijver},
  {Tarbell}, \& {Tomczyk}}]{2012SoPh..275..207S}
{Scherrer}, P.~H., {Schou}, J., {Bush}, R.~I., {et~al.} 2012, \solphys, 275,
  207

\bibitem[{{Snodgrass}(1983)}]{Snodgrass_1983}
{Snodgrass}, H.~B. 1983, \apj, 270, 288

\bibitem[{{Solanki} {et~al.}(2002){Solanki}, {Krivova}, {Sch{\"u}ssler}, \&
  {Fligge}}]{Solanki2002}
{Solanki}, S.~K., {Krivova}, N.~A., {Sch{\"u}ssler}, M., \& {Fligge}, M. 2002,
  \aap, 396, 1029

\bibitem[{Tlatov {et~al.}(2015)Tlatov, Tlatova, Vasil’eva, Pevtsov, \&
  Mursula}]{TLATOV2015835}
Tlatov, A.~G., Tlatova, K., Vasil’eva, V., Pevtsov, A., \& Mursula, K. 2015,
  Advances in Space Research, 55, 835, cosmic Magnetic Fields

\bibitem[{Tlatov {et~al.}(2014)Tlatov, Vasil’eva, Makarova,
  {et~al.}}]{Tlatov2014}
Tlatov, A.~G., Vasil’eva, V.~V., Makarova, V.~V., {et~al.} 2014, Solar
  Physics, 289, 1403

\bibitem[{{Upton} \& {Hathaway}(2014)}]{Upton2014}
{Upton}, L., \& {Hathaway}, D.~H. 2014, \apj, 780, 5

\bibitem[{{Upton} \& {Hathaway}(2018)}]{Upton2018}
{Upton}, L.~A., \& {Hathaway}, D.~H. 2018, \grl, 45, 8091

\bibitem[{{van Ballegooijen} {et~al.}(1998){van Ballegooijen}, {Cartledge}, \&
  {Priest}}]{Ballegooijen_1998}
{van Ballegooijen}, A.~A., {Cartledge}, N.~P., \& {Priest}, E.~R. 1998, \apj,
  501, 866

\bibitem[{{Vaquero}(2007)}]{Vaquero2007}
{Vaquero}, J.~M. 2007, Advances in Space Research, 40, 929

\bibitem[{{Vaquero} \& {Trigo}(2008)}]{Vaquero2008}
{Vaquero}, J.~M., \& {Trigo}, R.~M. 2008, \solphys, 250, 199

\bibitem[{{Virtanen} {et~al.}(2022){Virtanen}, {Pevtsov}, {Bertello}, \&
  {Mursula}}]{Virtanen2022}
{Virtanen}, I.~O.~I., {Pevtsov}, A.~A., {Bertello}, L., \& {Mursula}, K. 2022,
  \aap, 667, A168

\bibitem[{Wang {et~al.}(2014)Wang, Colaninno, Baranyi, \& Li}]{Wang_2015}
Wang, Y.-M., Colaninno, R.~C., Baranyi, T., \& Li, J. 2014, The Astrophysical
  Journal, 798, 50

\bibitem[{{Wang} \& {Sheeley}(1991)}]{Wang1991}
{Wang}, Y.~M., \& {Sheeley}, Jr., N.~R. 1991, \apj, 375, 761

\bibitem[{{Wang} {et~al.}(2021){Wang}, {Jiang}, \& {Wang}}]{Wang2021}
{Wang}, Z.-F., {Jiang}, J., \& {Wang}, J.-X. 2021, \aap, 650, A87

\bibitem[{{Watanabe} {et~al.}(2019){Watanabe}, {Imada}, {Iijima}, {Shiota}, \&
  {Miyoshi}}]{Watanabe2019}
{Watanabe}, Y., {Imada}, S., {Iijima}, H., {Shiota}, D., \& {Miyoshi}, Y. 2019,
  in AGU Fall Meeting Abstracts, Vol. 2019, SH43E--3387

\bibitem[{Whitbread {et~al.}(2017)Whitbread, Yeates, Muñoz-Jaramillo, \&
  Petrie}]{Whitbread_2017}
Whitbread, T., Yeates, A.~R., Muñoz-Jaramillo, A., \& Petrie, G. J.~D. 2017,
  Astronomy \& Astrophysics, 607, A76

\bibitem[{{Yeates}(2020)}]{Yeates2020}
{Yeates}, A.~R. 2020, \solphys, 295, 119

\bibitem[{Yeates {et~al.}(2015)Yeates, Baker, \& van
  Driel-Gesztelyi}]{Yeates2015}
Yeates, A.~R., Baker, D., \& van Driel-Gesztelyi, L. 2015, Solar Physics, 290,
  3189

\bibitem[{Yeates {et~al.}(2023)Yeates, Cheung, Jiang, {et~al.}}]{Yeates2023}
Yeates, A.~R., Cheung, M. C.~M., Jiang, J., {et~al.} 2023, Space Science
  Reviews, 219, 31

\bibitem[{{Zwaan}(1981)}]{Zwaan1981}
{Zwaan}, C. 1981, in NASA Special Publication, ed. S.~{Jordan}, Vol. 450,
  163--179

\end{thebibliography}

\end{document}